\documentclass[12pt]{article}

\usepackage{amsmath}
\usepackage{amssymb}
\usepackage{epsfig}
\usepackage{graphicx}
\usepackage{cite}
\usepackage{multirow}
\usepackage{longtable}
\usepackage{lscape}
\usepackage{bbm}
\usepackage{mathrsfs}

\allowdisplaybreaks[4] 

\newcommand{\capdef}{}
\newcommand{\mycaption}[2][\capdef]{\renewcommand{\capdef}{#2}%
        \caption[#1]{{\footnotesize #2}}}
\makeatletter
\renewcommand{\fnum@table}{\textbf{\tablename~\thetable}}
\renewcommand{\fnum@figure}{\textbf{\figurename~\thefigure}}
\makeatother

\newcounter{myenumi}

\renewcommand{\themyenumi}{\roman{myenumi}}
{\end{list}}

\newlength{\myem}
\newcounter{mysubequation}[equation]

\makeatletter
\renewcommand{\section}{\@startsection{section}{1}{0em}{-\baselineskip}%
{\baselineskip}{\normalfont\large\bfseries}}
\renewcommand{\subsection}%
{\@startsection{subsection}{2}{0em}{-0.7\baselineskip}%
{0.7\baselineskip}{\normalfont\bfseries}}
\makeatother

\newcommand{\bi}{\begin{itemize}}
\newcommand{\ei}{\end{itemize}}

\newcommand{\be}{\begin{equation}}
\newcommand{\ee}{\end{equation}}
\newcommand{\bea}{\begin{eqnarray}}
\newcommand{\eea}{\end{eqnarray}}

\newcommand{\ldm}{\Delta m_{31}^2}
\newcommand{\sdm}{\Delta m_{21}^2}
\newcommand{\deltacp}{\delta_{\mathrm{CP}}}
\newcommand{\stheta}{\sin^2 2 \theta_{13}}

\newcommand{\ie}{{\it i.e.}}

\newcommand{\eg}{{\it e.g.}}

\newcommand{\cf}{{\it cf.}}

\newcommand{\etc}{{\it etc.}}
\newcommand{\eq}{Eq.}

\newcommand{\fig}{Fig.}

\newcommand{\Ref}{Ref.}
\newcommand{\Refs}{Refs.}
\newcommand{\Sec}{Sec.}

\newcommand{\App}{Appendix}

\newcommand{\Tab}{Table}

\newcommand{\equ}[1]{\eq~(\ref{equ:#1})}
\newcommand{\figu}[1]{\fig~\ref{fig:#1}}

\begin{document}

\begin{titlepage}

\renewcommand{\thefootnote}{\alph{footnote}}

\vspace*{-3.cm}
\begin{flushright}
EURONU-WP6-09-02
\end{flushright}

\vspace*{0.5cm}

\renewcommand{\thefootnote}{\fnsymbol{footnote}}
\setcounter{footnote}{-1}

{\begin{center}
{\large\bf
Physics with near detectors at a neutrino factory
} \end{center}}
\renewcommand{\thefootnote}{\alph{footnote}}

\vspace*{.8cm}
\vspace*{.3cm}
{\begin{center} {\large{\sc
 		Jian~Tang\footnote[1]{\makebox[1.cm]{Email:}
                jtang@physik.uni-wuerzburg.de},
                Walter~Winter\footnote[2]{\makebox[1.cm]{Email:}
                winter@physik.uni-wuerzburg.de}
                }}
\end{center}}
\vspace*{0cm}
{\it
\begin{center}

\footnotemark[1]${}^,$\footnotemark[2]
       Institut f{\"u}r Theoretische Physik und Astrophysik, Universit{\"a}t W{\"u}rzburg, \\
       D-97074 W{\"u}rzburg, Germany

\end{center}}

\vspace*{1.5cm}

{\Large \bf
\begin{center} Abstract \end{center}  }

We discuss the impact of near detectors at a neutrino factory both on standard oscillation and non-standard interaction measurements. Our systematics treatment includes cross section errors, flux errors, and background uncertainties, and our near detector fluxes include the geometry of the neutrino source and the detector. Instead of a specific detector concept,
we introduce qualitatively different classes of near detectors with different characteristics, such as near detectors catching the whole neutrino flux (near detector limit) versus near detectors observing a spectrum similar to that of the far detector (far detector limit). We include the low energy neutrino factory in the discussion.
We illustrate for which measurements near detectors are required, discuss how many are needed, and what the role of the flux monitoring is. For instance, we demonstrate that near detectors are mandatory for the leading atmospheric parameter measurements if the neutrino factory has only one baseline, whereas systematical errors partially cancel if the neutrino factory complex includes the magic baseline. 
Finally, near detectors with $\nu_\tau$ detection are shown to be useful for non-standard interactions.

\vspace*{.5cm}

\end{titlepage}

\newpage

\renewcommand{\thefootnote}{\arabic{footnote}}
\setcounter{footnote}{0}

\section{Introduction}

In neutrino physics, three-flavor oscillations have been successful to explain all 
relevant neutrino data, see, \eg, \Ref~\cite{GonzalezGarcia:2007ib}.
In particular, the solar and atmospheric oscillation parameters have been
measured with very high precisions, and the reactor mixing angle $\theta_{13}$
has been strongly constrained. Future long-baseline and reactor neutrino 
experiments will test this small angle
further, and be sensitive to leptonic CP violation and the neutrino mass
hierarchy (see \Ref~\cite{Bandyopadhyay:2007kx} and references therein). The ultimate
high precision instrument for these purposes might be a neutrino factory~\cite{Geer:1998iz,DeRujula:1998hd,Barger:1999fs,Cervera:2000kp}.
Using different baselines and oscillation channels, it can basically disentangle
all of the remaining oscillation parameters~\cite{Donini:2002rm,Autiero:2003fu,Huber:2003ak,Huber:2006wb} 
in spite of the presence of intrinsic correlations and degeneracies~\cite{Fogli:1996pv,Cervera:2000kp,Minakata:2001qm,Huber:2002mx}.
Furthermore, a neutrino factory and other future neutrino oscillation experiments will
be sensitive to new physics, such as so-called non-standard interactions; see, \eg, \Refs~\cite{Bueno:2000jy,Huber:2001de,Huber:2001zw,Gonzalez-Garcia:2001mp,Ota:2001pw,Gago:2001xg,Campanelli:2002cc,Ota:2002na,Huber:2002bi,Hattori:2002uw,Garbutt:2003ih,Blennow:2005qj,Friedland:2006pi,Kitazawa:2006iq,Honda:2006gv,Adhikari:2006uj,Blennow:2007pu,Kopp:2007mi,Ribeiro:2007ud,Kopp:2007ne,Ribeiro:2007jq,EstebanPretel:2008qi,Blennow:2008ym,Malinsky:2008qn,Ohlsson:2008gx}.

The design of a neutrino factory has been put forward and discussed in international studies, such as in \Refs~\cite{Albright:2000xi, Apollonio:2002en,Albright:2004iw,Bandyopadhyay:2007kx}. Especially the most recent study, the International Neutrino Factory and Superbeam Scoping Study~\cite{Bandyopadhyay:2007kx,Abe:2007bi,Berg:2008xx}, has laid the foundations for the currently ongoing Design Study for the Neutrino Factory (IDS-NF)~\cite{ids}. This
initiative from about 2007 to 2012 is aiming to present a design report, schedule, cost estimate, and risk assessment for a neutrino factory. It defines a baseline setup of a high energy neutrino factory with two baselines
$L_1 \simeq 4 \, 000 \, \mathrm{km}$ and $L_2 \simeq 7 \, 500 \, \mathrm{km}$ (the ``magic'' baseline) operated by two racetrack-shaped storage rings, where the muon energy is 25~GeV (for optimization questions, see \Refs~\cite{Huber:2006wb,Gandhi:2006gu}). There are no near detector specifications yet, and the systematics treatment is done in an effective way by signal and background normalization errors uncorrelated among all channels and detectors. Therefore, there are a number of questions currently raised within the IDS-NF:
\begin{itemize}
\item
 Study of the potential of near detectors to cancel systematical errors.
\item
 Study of the characteristics of the near detectors, such as technology, number, \etc.
\item
 Study of the use of the near detectors for searches of new physics.
\end{itemize}
In this work, we address several of these questions using a more refined systematics treatment including cross section errors, flux errors, and background uncertainties. Similar systematics studies using the pull method for physics studied have so far been especially used in the context of reactor experiments, see, \eg, \Refs~\cite{Huber:2003pm,Huber:2006vr}.
A systematics discussion of superbeams can be found in \Ref~\cite{Huber:2007em}. We also include a low energy version of the neutrino factory in the discussion~\cite{Geer:2007kn,Huber:2007uj,Bross:2007ts}, which has been recently proposed to reduce the accelerator cost in the case of large $\theta_{13}$.

For the near detectors at a neutrino factory, a summary of options can be found in \Ref~\cite{Abe:2007bi}. The near detectors are, for example, supposed to measure the neutrino flux, the neutrino beam angle and its divergence, the neutrino energy, and the neutrino cross sections. 
For the technology, liquid argon time projection chambers, conventional scintillators, scintillating fiber trackers, gas time projection chambers, silicon detectors, and emulsion detectors are mentioned as possible options. We will not specify a particular detector technology, but instead postulate that the near detectors can at least measure the $\nu_\mu$ (or $\bar\nu_\mu$) event rates with (at least) the same energy resolution as the far detectors. Since  close to the neutrino source the geometry of both the source and the detector become important, we introduce qualitatively different classes of near detectors with different characteristics, such as near detectors catching the whole neutrino flux (near detector limit) versus near detectors observing a spectrum similar to the far detector (far detector limit). We do not include any additional properties, such as beam angle measurements, in our treatment. The main purpose of this study is not an accurate description of the near detectors with all their properties, but an estimate of when near detectors are important from the physics point of view.

The size of these multi-purpose detectors will typically be small, such as $\mathcal{O}(100)$~kg, and the charged current event rates extremely high compared to that in the far detector. To a first approximation, the near detectors measure the product of flux and cross sections. However, using different types of charged current interactions (such as the purely leptonic inverse muon decays with a different threshold of about 10.9~GeV), one can, in principle, disentangle the fluxes and cross sections in the near detector. Since these interactions have relatively small cross sections, larger near detectors may be favorable at the end. We do not enter this level of detail, but only assume that the near detectors measure the inclusive charged current cross sections. We choose different sizes for the near detectors, from small (200~kg), over ``typical'' (of the size of the SciBOONE, MINER$\nu$A, NOMAD, or the MINOS near detectors) to a large (hypothetical) detector to capture the whole beam. For the flux knowledge, we make rather conservative assumptions. For example, the total number of muons circulating in a storage ring may be inferred from a Beam Current Transformer, such that the flux is known to the level of $10^{-3}$. However, if the muon beam divergence is too large (and not monitored), additional uncertainties enter the flux. Therefore, we start with a much more conservative assumption (2.5\%), and illustrate the improved flux knowledge from different interaction types in the near detectors and beam monitoring separately.

Note that the purpose if this work is to provide information for the development of near detectors, in particular, for which physics measurements near detectors are required. Of course, systematics is closely connected to that. 
Our study is to be interpreted as a guideline for the minimal requirements to near detectors and systematics in order to obtain to the physics wanted.

This study is organized as follows: We discuss the neutrino factory flux in \Sec~\ref{sec:flux}. Then we introduce our near detectors in \Sec~\ref{sec:nd}, and compute the corresponding spectra. In \Sec~\ref{sec:syst} we describe our systematics treatment and simulation methods, where details can be found in \App~\ref{app:syst}. In \Sec~\ref{sec:so} we show the physics results for the high energy neutrino factory and in \Sec~\ref{sec:solow} for the low energy neutrino factory, both for standard oscillations. Furthermore, we discuss in \Sec~\ref{sec:nsi} the non-standard interactions in the context of the high energy neutrino factory, before we summarize in \Sec~\ref{sec:summary}.

\section{Neutrino factory flux}
\label{sec:flux}

\begin{figure}[t]
\begin{center}
$\begin{array}{ll}
 \includegraphics[width=0.48\textwidth]{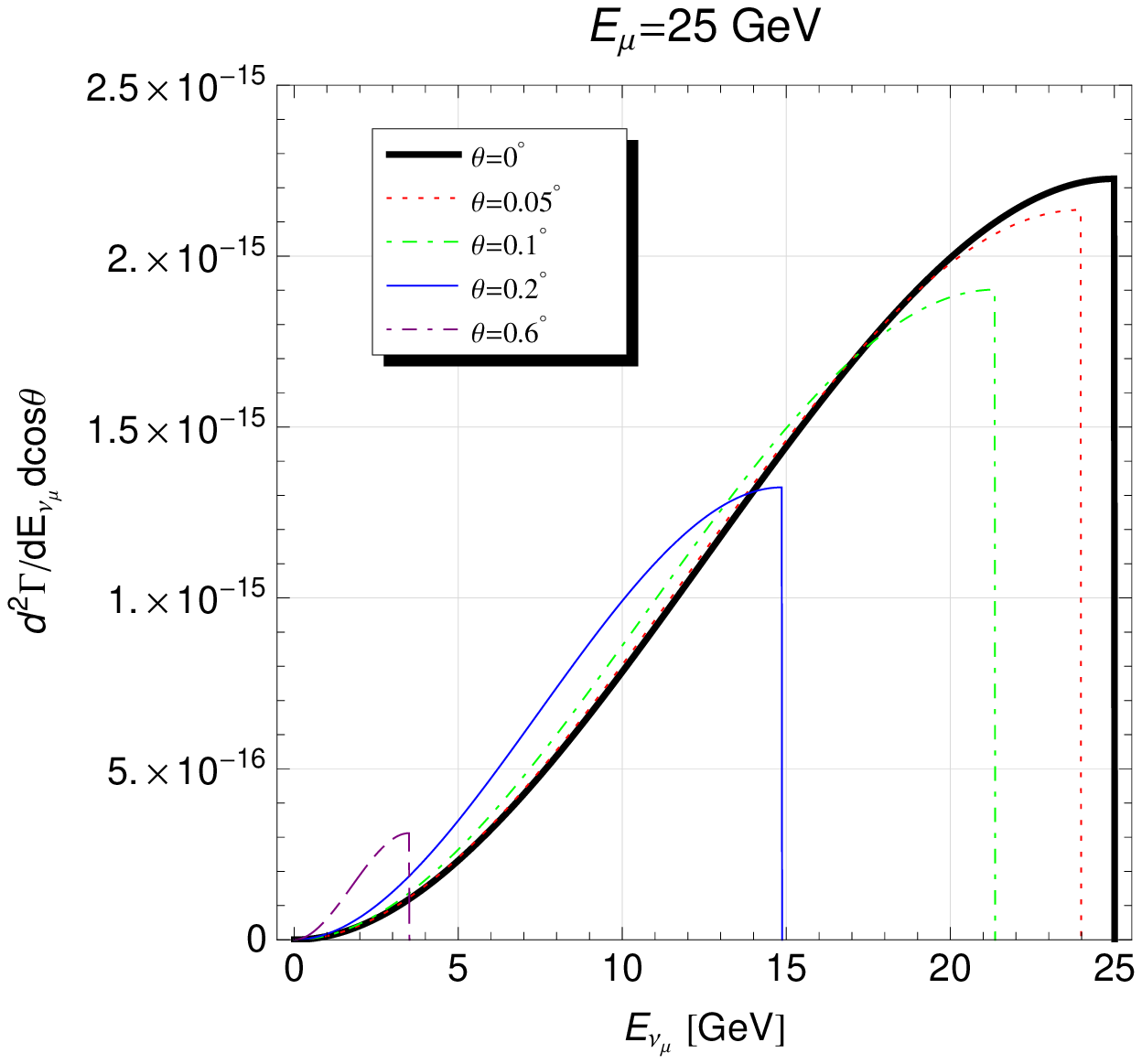}
&\includegraphics[width=0.48\textwidth]{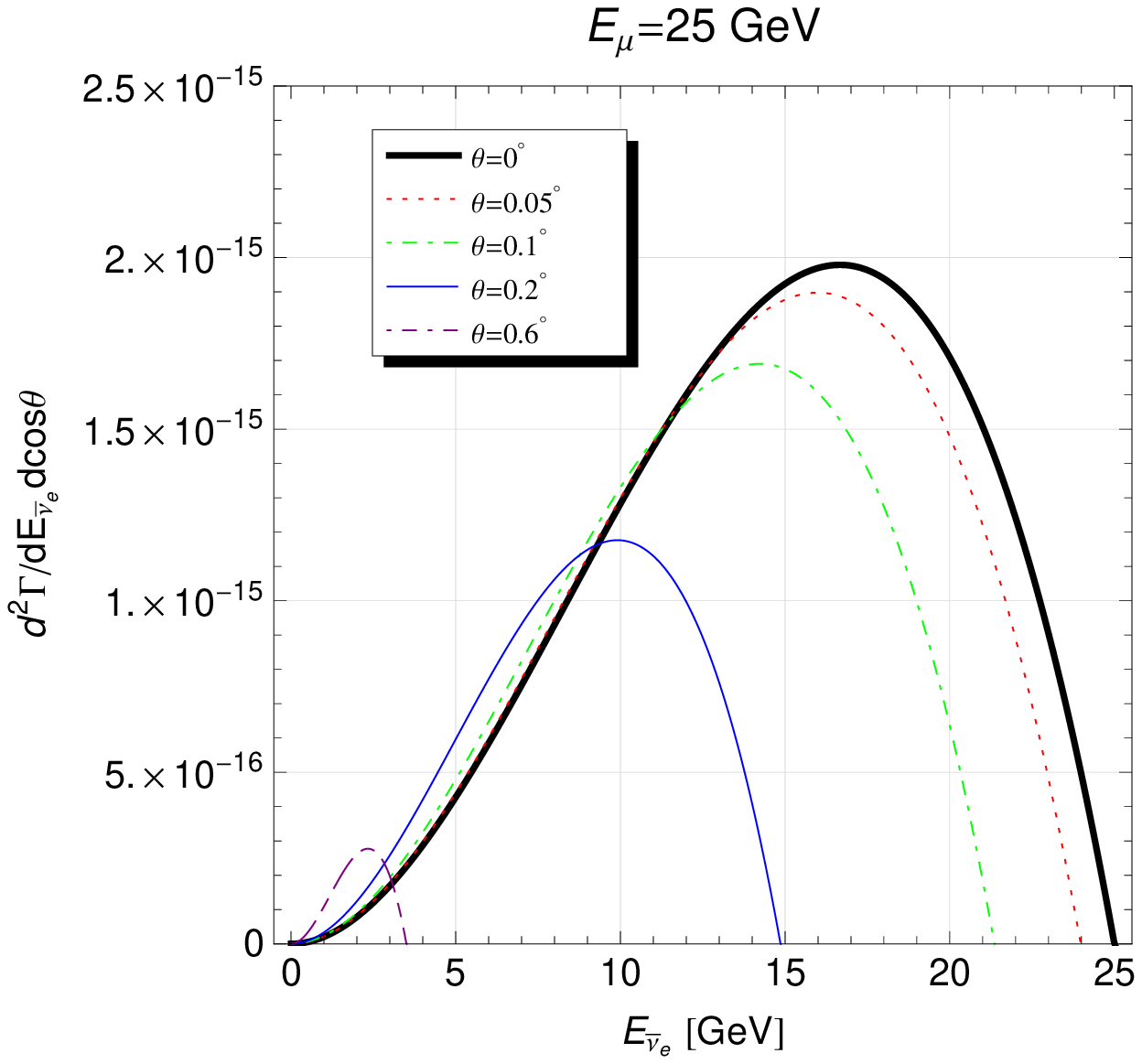}
\end{array}$
\end{center}
\mycaption{The unpolarized differential neutrino spectra in the laboratory frame 
for $\nu_\mu$ (left panel) and $\nu_e$ (right panel), where
$E_{\mu}=25$~GeV and different off-axis angles are used (see legends, in degrees).}
\label{fig:oaspectra}
\end{figure}

The (double) differential decay rates of an unpolarized muon in the laboratory frame are given by (see, \eg, \Refs~\cite{Geer:1998iz,DeRujula:1998hd,Barger:1999fs,Cervera:2000kp})
\begin{eqnarray}
 \frac{d^2\Gamma}{dE_{\nu_\mu}d\cos\theta}&=&\frac{G_F^2 m_\mu}{24 \pi^3} \gamma (1- \beta \cos\theta) E_{\nu_\mu}^2 \left[ 3 m_\mu -4\gamma E_{\nu_\mu}(1- \beta \cos\theta)\right]\,,
\label{equ:specmu}
\\
 \frac{d^2\Gamma}{dE_{\nu_e} d\cos\theta}&=&\frac{G_F^2 m_\mu}{4\pi^3}\gamma (1- \beta \cos\theta) E_{\nu_e}^2 \left[m_\mu - 2\gamma E_{\nu_e}(1-\beta \cos\theta)\right]\,.
\label{equ:spece}
\end{eqnarray}
Here $\nu_\alpha$ ($\alpha \in \{e, \mu\}$) stands for both $\nu_\alpha$ and $\bar\nu_\alpha$, $\gamma= E_\mu/m_\mu = 1/\sqrt{1 - \beta^2}$ is the boost factor, and $m_\mu$ is the muon rest mass. The angle $\theta$ is the angle between the travel direction of the muon and the observer (in the laboratory frame, at the decay point), which we henceforth call {\bf off-axis angle}. 
Note that in a racetrack-shaped storage ring, any muon polarization averages out with a high precision~\cite{Abe:2007bi}, which means that we can use the above unpolarized spectra. In addition, note that the theoretical knowledge on these fluxes (the calculated spectrum can be trusted at the level to $10^{-3}$) is expected to be more precise than needed for this study.

Very importantly, the off-axis angle has to satisfy the flavor-independent constraint 
\begin{equation}
\theta \le \theta_{\mathrm{cut}}   \quad \mathrm{with} \quad \theta_{\mathrm{cut}}(E_\nu) \simeq \frac{1}{\gamma} \sqrt{\frac{E_\mu}{E_\nu} -1 } \, ,
\label{equ:thetacut}
\end{equation}
which is a relationship depending on the neutrino energy. Conversely, the inverse of this relationship limits the neutrino energy observable at a certain off-axis angle, \ie, the flux is zero for $E_\nu> E_{\nu,\mathrm{cut}}(\theta) \simeq E_\mu/(1+(\gamma \theta)^2)$, which means that the off-axis angle $\theta$ determines the maximal available neutrino energy. It needs to be included in any integration over off-axis angle (such as over the detector's surface area) or the energy. We illustrate the relationship in \figu{oaspectra}, where especially the $\nu_\mu$ spectrum faces a sharp cutoff depending on the off-axis angle. At $E_\nu> E_{\nu,\mathrm{cut}}(\theta)$, the flux is zero. From \equ{thetacut}, for $E_\nu = E_\mu$, we have $\theta_{\mathrm{cut}} \simeq 0$, which means that the high energy part of the beam is going in the forward direction. For $E_\nu = E_\mu/2$, we have $\theta_{\mathrm{cut}} \simeq 1/\gamma$, which is often referred to as the opening angle of the beam (we will use a different, energy independent definition below). For small neutrino energies, the beam obviously becomes wider than that for high energies, which corresponds to the relative enhancement of the off-axis fluxes at low energies in \figu{oaspectra}.

\begin{figure}[t]
\centering
 \includegraphics[width=0.5\textwidth]{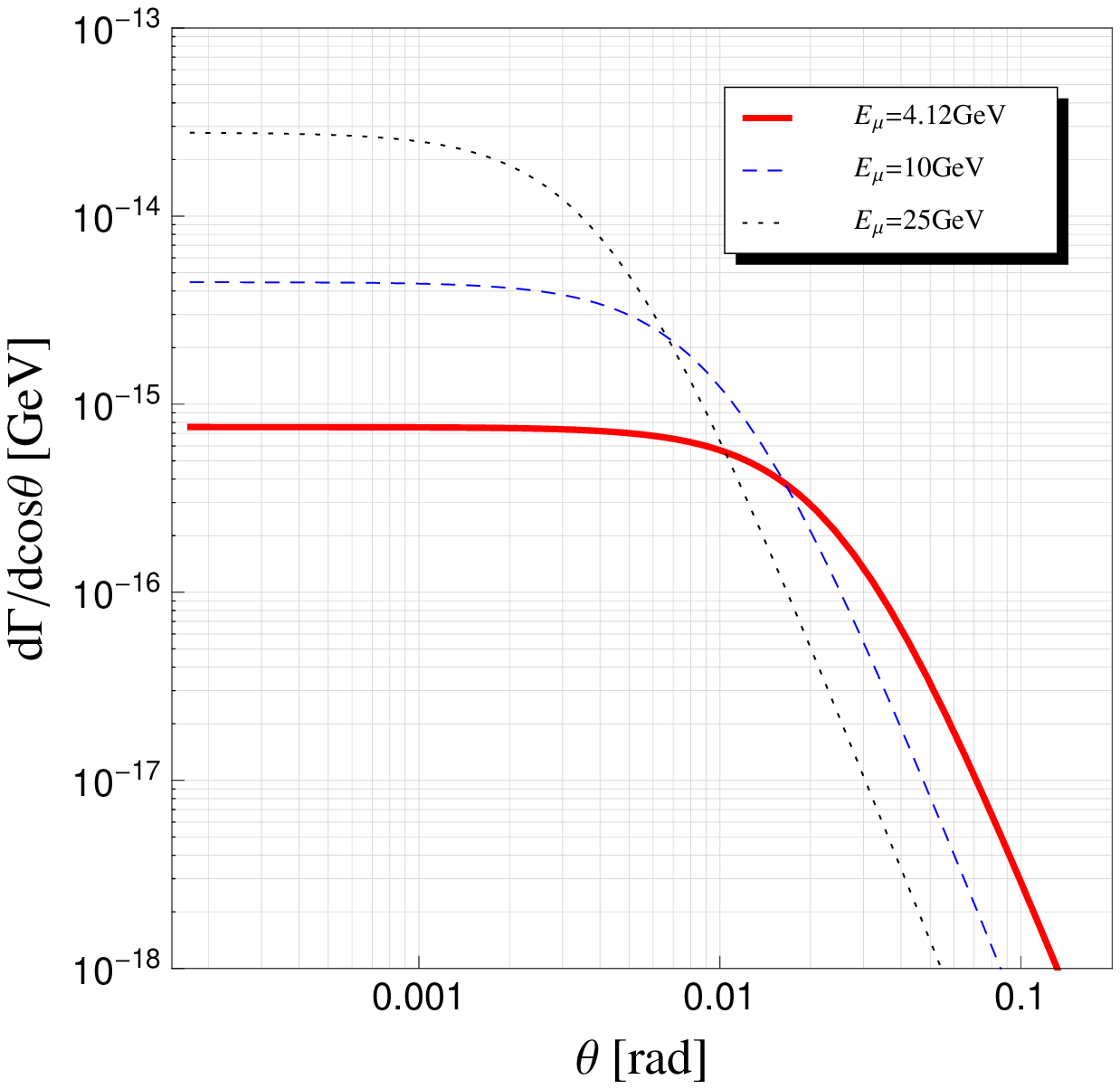}
\mycaption{The unpolarized single differential neutrino spectrum integrated over energy as a function of the off-axis angle $\theta$ (in radians) for different values of $E_\mu$.}
\label{fig:lab-Ecut}
\end{figure}

For the following discussion, it turns out to integrate out the energy in order to obtain the single differential decay rate (observing \equ{thetacut})
\begin{equation}
 \frac{d\Gamma}{d\cos\theta}=\frac{G_F^2 m_\mu^5}{384 \pi^3}\frac{1}{[\gamma(1- \beta \cos\theta)]^2} \, ,
\label{equ:gammatheta}
\end{equation}
which we show in \figu{lab-Ecut}. Obviously, the energy-integrated decay rate always peaks at $\theta=0$, and decreases monotoneously with the off-axis angle.
We define the (neutrino) {\bf beam divergence} $\hat\theta$ by the equation
\begin{equation}
 \frac{1}{\Gamma_0} \int\limits_0^{\hat\theta} \frac{d\Gamma}{d\cos\theta} \sin \theta d\theta=\frac{1}{2} \int\limits_0^{\hat\theta} \frac{1}{\gamma^2(1-\beta\cos\theta)^2} \sin \theta d \theta = \wedge\, ,
\label{equ:div}
\end{equation}
where $\Gamma_0$ is the total decay rate (integrated over angle and energy)  $\Gamma_0=\frac{G_F^2m_\mu^5}{192\pi^3}$. This means that the fraction $\wedge$ of the total flux is contained in the cone limited by the angle $\hat\theta$. 
The beam divergence can be used to define a near detector which captures the total flux of the beam. Note that we imply that the muon beam divergence be much smaller than the neutrino beam divergence, since otherwise a sizable amount of events will be lost~\cite{Abe:2007bi}.

Similarly, we define the {\bf beam opening angle} $\tilde \theta$ as 
\begin{equation}
\frac{d \Gamma}{d \cos \theta} |_{\theta=\tilde \theta} = \wedge \cdot \frac{d \Gamma}{d \cos \theta} |_{\theta=0} \, ,
\label{equ:op}
\end{equation}
which quantifies the angle over which the flux stays almost constant. It can be used to define a near detector
which observes a flux similar to that of a far detector, for which the flux at $\theta=0$ is used. Note that this particular definition of the opening angle does not depend on the neutrino energy, which we have integrated out in \equ{gammatheta}, whereas \equ{thetacut} depends on the neutrino energy.
\begin{table}[t]
\begin{center}
\begin{tabular}{lrrrr}
\hline
 & \multicolumn{2}{c}{Beam divergence $\hat\theta$} & \multicolumn{2}{c}{Beam opening angle $\tilde \theta$} \\
 & $\wedge=0.90$ & $\wedge=0.99$ & $\wedge=0.90$ & $\wedge=0.99$ \\
\hline
$E_\mu = 25 \, \mathrm{GeV}$ & 0.0127 & 0.0420 & 0.000983 & 0.000300 \\
$E_\mu = 4.12 \, \mathrm{GeV}$ & 0.0769 & 0.2538 & 0.005966 & 0.001821 \\
\hline
\end{tabular}
\end{center}
\mycaption{\label{tab:div}The beam divergence $\hat\theta$ (as defined in \equ{div}) and the
beam opening angle $\tilde \theta$ (as defined in \equ{op}) for two different muon energies and
two different fractions $\wedge$ in radians. These numbers are obtained from \figu{lab-Ecut}.}
\end{table}
We show $\hat\theta$ and $\tilde\theta$ in \Tab~\ref{tab:div} for two different muon energies. Obviously, the larger $\wedge$ becomes, the smaller is $\tilde \theta$, and the larger is $\hat \theta$.
The actual diameter of the beam for these two quantities is finally obtained as 
\begin{equation}
D \simeq 2 \times L \times \theta \, 
\label{equ:dia}
\end{equation}
where $L$ (baseline) is the distance between production point and observer.

\section{Definition and spectra of the near detectors}
\label{sec:nd}

Here we define our near detectors and describe the computations of the spectra.

\subsection{Near detector definitions}

\begin{figure}[t]
\includegraphics[width=\textwidth]{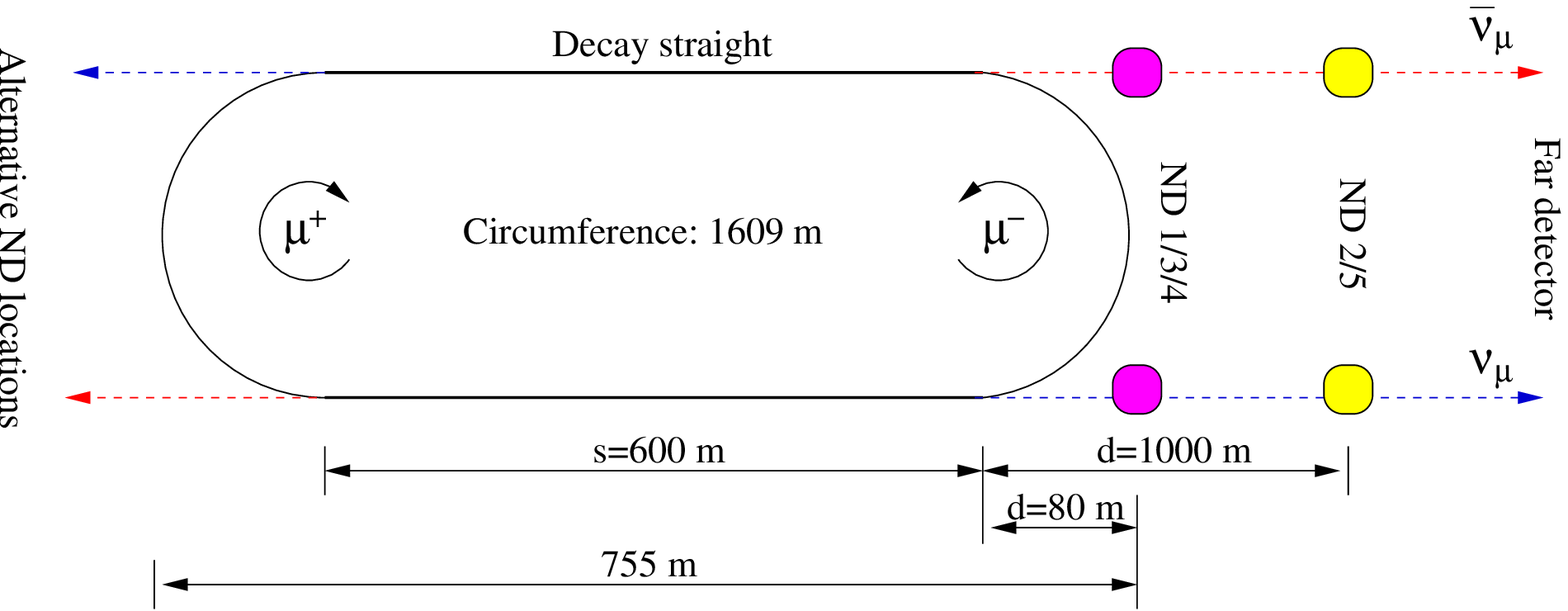}
\mycaption{\label{fig:ring}Geometry of the muon storage ring and possible near detector (ND) locations (not to scale). The baseline $L$ is the distance between production point and near detector, \ie, $d \le L \le d+s$.}
\end{figure}

As indicated in the introduction, we are interested in the conceptually different detectors from the flux point of view, where the sole purpose is the flux $\times$ cross section measurement. 
 We use the racetrack-shaped storage 
ring geometry from IDS-NF~\cite{ids} designed for a high energy neutrino factory. This geometry together with possible near detector locations (to be defined later) is illustrated in \figu{ring}.
In general, for a near detector, the far distance and point source approximations
\begin{equation}
\frac{d^2\Gamma}{dE_{\nu_\alpha}d\cos\theta} (\cos \theta) \simeq \frac{d^2\Gamma}{dE_{\nu_\alpha}d\cos\theta}|_{\theta = 0} \, , \quad
L \gg s \quad \mathrm{(Size \, \,  of \, \, source)}
\label{equ:ps}
\end{equation}
do not hold, and the geometry cannot be neglected.

Our near detector definitions will be based on the following assumptions or observations:
\begin{itemize}
\item
 We operate the near detectors on-axis, because for an off-axis operation, the event rates at high energies
will be suppressed according to Eqs.~(\ref{equ:specmu}) and~(\ref{equ:spece}).
\item
 The only purpose of the near detectors is the $\nu_\mu$ (from $\mu^-$ decays) and
the $\bar\nu_\mu$ (from $\mu^+$ decays) event rate measurement using the inclusive charged current cross sections. Since the muons and anti-muons are assumed to circulate in different directions in the storage ring (\cf, \figu{ring}), 
two near detectors are required.
\item
 We do not measure the $\nu_e$ and $\bar\nu_e$ event rates, since we do not need the
corresponding cross sections in the neutrino factory far detector.\footnote{In fact, not even the dominating backgrounds (charge mis-identification and neutral currents) depend on the electron neutrino cross sections.} Therefore, we avoid to complicate the
cross section discussion (such as to relate the muon neutrino and electron neutrino cross sections).
\item
 We use the same characteristics as in the far detectors, such as energy resolution and binning, for the sake of simplicity.
\item
 We do not extrapolate the backgrounds from the near to the far detectors, but instead use relatively large background uncertainties uncorrelated among all channels.
\item
 We assume the fiducial volume to be cylindrical. In addition, we assume that the detector only sees muon decays
from the decay straight (denoted by $s$ in \figu{ring}).\footnote{The contributions from the curved sections can be estimated to be at the few percent level and can be easily computed once the final geometry is known.}
\item
 We do not include systematical errors potentially uncorrelated among the detectors, such as
fiducial volume errors and energy calibration errors. These additional systematical errors may
affect the final results, but they cannot be improved upon by near detectors.
\end{itemize}
Note that we do not necessarily require a magnetic field or charge identification 
since our distance is too short for the neutrinos to oscillate, but we need sufficient
flavor identification at the level of (at least) the flux normalization error. 
Alternatively, the different flavors can be distinguished by the different charges,
but this may be technically more challenging.

\begin{table}[t]
\begin{center}
\begin{tabular}{lrrrrr}
\hline 
Parameter & ND1 & ND2 & ND3 & ND4 & ND5   \\
\hline
Diameter $D$ & 17~m & 4~m & 4~m & 0.32~m & 6.8~m  \\
Distance $d$ & 80~m & 1000~m & 80~m & 80~m &  1000~m \\
Mass & 450~t & 25~t & 25~t & 0.2~t & 2000~t \\
\hline
\end{tabular}
\end{center}
\mycaption{\label{tab:nd} Definition of our near detector fiducial (active) volumes in terms of diameter, distance $d$ to the end of the decay straight (\cf, \figu{ring}), and mass. The fiducial volumes are assumed to be cylindrical. If the density is about $1 \, \mathrm{g \cdot cm^{-3}}$ (such as for a liquid scintillator), the active detectors will be about 2~m long for ND1 to ND4. ND5 is assumed to be OPERA-sized (with a cylindrical shape for the sake of simplicity). For the low energy neutrino factory, only ND3 is used. ND4 is a down-scale (tabletop) version of ND2 with the same ratio $D/L$. The ND mass of 25~t is at the upper limit of currently used near detectors.}
\end{table}

From the conceptual point of view, we postulate that the geometry determines the limiting cases: 
\begin{description}
\item[Near detector limit] In this case, the neutrino beam divergence given by \equ{div}, applied to \equ{dia}, is smaller than the detector diameter for the farthest decay point of the decay straight $L=d+s$, \ie, the full flux (integrated over the angle) is seen by the detector from any decay point in the straight.
\item[Far detector limit] In this case, the beam diameter given by the opening angle in \equ{op} applied to \equ{dia} is of the order of the detector diameter at the nearest decay point $L=d$, \ie, the far distance approximation in \equ{ps} (first condition) is approximately fulfilled for any decay point in the straight.
This implies that the beam spectrum will be similar to that of a far detector.
\end{description}
Our near detector parameters are shown in \Tab~\ref{tab:nd}. Note that the active volumes are much smaller than the actual detectors, because the secondary muons have to be stopped. There we define a hypothetical ND1 as a detector operating in the near detector limit, and a ND2 as a detector operating close to the far detector limit. For example,  we find from \equ{dia} using $\hat\theta$ in \Tab~\ref{tab:div} ($\wedge=0.9$) that the beam diameter is about 17~m for $L=s+d= 680 \, \mathrm{m}$, which explains the large diameter of ND1 to catch the whole flux.
ND3 is an intermediate case between the near and far limits, as we will demonstrate later.
The size of ND2 and ND3 is similar to conventional near detectors, such as SciBOONE, MINER$\nu$A, NOMAD or the MINOS near detector. ND4 is a smaller version of ND2 with the same ratio between detector diameter and distance $d$. If the source was a point source (\ie, the straight would be a point), the event rate would be exactly the same as in ND2. Finally, ND5 is an OPERA-sized near detector, which we will only use for non-standard physics tests. For a low energy, no storage ring is specified yet. Therefore, we assume the same geometry as in \figu{ring} (although a smaller storage ring could be sufficient). Since the beam is much wider because of the smaller boost factor (\cf, \Tab~\ref{tab:div}), ND3 will already perform similar to a far detector. Therefore, we only use ND3 for this option.
Of course, one can always up-scale or down-scale the discussed detectors (to re-scale the event rate), or change the distance to the source. However, from this qualitative discussion, we have covered all relevant cases allowed by the source geometry.  In fact, ND4 may come closest to a realistic near detector.

\subsection{Near detector fluxes}

In order to compute the fluxes for the near detectors, let us first of look at the number of muon neutrino events produced in a near detector. From the very first principles, it is given by
\begin{equation}
 \frac{dN}{dE}= \frac{\sigma}{A} \, \frac{d N_{\mathrm{Beam}}}{dE} \, N_\mathrm{Det} = 
\sigma   \frac{1}{A_{\mathrm{Det}}} \int\limits_{A_{\mathrm{Det}}} \frac{d^2\Phi}{dE dA} dA \, \frac{M_{\mathrm{Det}}}{m_N} \, ,
\label{equ:ne}
\end{equation}
where $N_{\mathrm{Beam}}$ is the number of neutrinos in the beam (within the detector), $N_\mathrm{Det}$ is the number of target nucleons in the detector, $\sigma$ is the cross section per nucleon, and $m_N$ is the nucleon mass. The flux $\Phi$ is, for a cylindrical detector, related to the decay rate $\Gamma$ by
\begin{equation}
 \frac{d^2\Phi}{dE dA} = \frac{n_\mu}{\Gamma_0} \frac{d^2 \Gamma}{dE d \cos\theta} \, \frac{1}{2 \pi L^2}
\label{equ:phigamma}
\end{equation}
with $n_\mu$ the number of useful muon decays, $\Gamma_0$ total decay rate, and $L$ the baseline.
Applying the far distance and point source approximations \equ{ps} in Eqs.~(\ref{equ:ne}) and~(\ref{equ:phigamma}), we have for the ``point source'' (PS) event rate 
\begin{equation}
 \frac{dN_{\mathrm{PS}}}{dE} \simeq \frac{n_\mu}{\Gamma_0} \, \sigma \, \frac{1}{A_{\mathrm{Det}}} \, \int\limits_{0}^{\theta_{\mathrm{max}}} \frac{d^2\Gamma}{dE d\cos\theta}|_{\theta=0}  \, \sin \theta \, d \theta \, \frac{M_{\mathrm{Det}}}{m_N} \simeq 
\frac{n_\mu}{\Gamma_0} \, \sigma  \,  \frac{d^2\Gamma}{dE d\cos\theta}|_{\theta=0} \, \frac{M_{\mathrm{Det}}}{m_N}\,  \frac{1}{2 \pi L^2} \, ,
\label{equ:nps}
\end{equation}
where $\theta_{\mathrm{max}} \simeq D/(2L)$ is related to the detector diameter $D$. Therefore, only the on-axis flux is
needed, the event rate increases linearly with detector mass, and drops as $1/L^2$. Such a flux is typically used for long-baseline experiment simulations, such as in GLoBES~\cite{Huber:2004ka,Huber:2007ji}.

In our case, we cannot use these approximations for the near detectors. Therefore, we proceed in two steps: First, we fix the production point and integrate over the surface area of the detector. And second, we integrate over the decay straight.
For a fixed production point in a distance $L$ from the detector, we obtain similar to \equ{nps}
\begin{equation}
 \frac{dN}{dE} \simeq \frac{n_\mu}{\Gamma_0} \, \sigma \, \frac{1}{A_{\mathrm{Det}}} \, \int\limits_{0}^{\frac{D}{2L}} \frac{d^2\Gamma}{dE d\cos\theta}  \, \sin \theta \, d \theta \, \frac{M_{\mathrm{Det}}}{m_N} = \frac{dN_{\mathrm{PS}}}{dE} \, \frac{A_{\mathrm{eff}}}{A_{\mathrm{Det}}}  \equiv
\frac{dN_{\mathrm{PS}}}{dE} \, \varepsilon(E,L)
\label{equ:n}
\end{equation}
with the effective area $A_{\mathrm{eff}}$ and the efficiency ratio $\varepsilon(E,L)$ determined by
\begin{equation} A_{\mathrm{eff}} = \frac{2 \pi L^2}{\frac{d^2\Gamma}{dE d\cos\theta}|_{\theta=0} } \, \int\limits_{0}^{\frac{D}{2L}} \frac{d^2\Gamma}{dE d\cos\theta}  \, \sin \theta \, d \theta 
\quad \mathrm{and} \quad \varepsilon(E,L)= \frac{A_{\mathrm{eff}}}{A_{\mathrm{Det}}} \,  ,
\label{equ:aeff}
\end{equation}
respectively.
Therefore, the event rate in an arbitrary near detector can be related to the point source event rate in \equ{nps} with the same $L$. Then the efficiency ratio $\varepsilon$ describes what fraction of the beam is captured compared to the on-axis flux. Note that the event rate is highest on-axis for high neutrino energies and off-axis for low neutrino energies, which means that $\varepsilon$ can be larger than one for low neutrino energies, because the detector then captures some of the low energy part which is not present in the point source approximation (\cf, \figu{oaspectra}: for the off-axis spectra, the low energy part is enhanced). From \equ{aeff}, one can read off that $A_{\mathrm{eff}}$ approaches $A_{\mathrm{Det}}$, or $\varepsilon$ goes to unity, if the far distance approximation in \equ{ps} is applied.

As the next step, we average \equ{n} over the decay straight assuming that the probability for muon decay is the same everywhere in the straight (\cf, \figu{ring}):
\begin{equation}
\frac{dN_{\mathrm{avg}}}{dE} = \frac{1}{s} \int\limits_{d}^{d+s}\frac{dN}{dE} dL = \frac{1}{s} \int\limits_{d}^{d+s}\frac{dN_{\mathrm{PS}}(L,E)}{dE} \, \varepsilon(L,E) dL \, .
\end{equation}
From \equ{nps}, we find that $\frac{dN_{\mathrm{PS}}(L,E)}{dE} \propto 1/L^2$. Therefore, we can pull out $\frac{dN_{\mathrm{PS}}(L,E)}{dE}$ from the integral in order to obtain
\begin{equation}
\frac{dN_{\mathrm{avg}}}{dE} = \frac{dN_{\mathrm{PS}}(L_{\mathrm{eff}},E)}{dE} \frac{L_{\mathrm{eff}}^2}{s} \int\limits_{d}^{d+s} \frac{ \varepsilon(L,E)}{L^2} dL  =  \frac{dN_{\mathrm{PS}}(L_{\mathrm{eff}},E)}{dE}
 \, \hat\varepsilon(E) \label{equ:eavg}
\end{equation}
with the average efficiency ratio
\begin{equation}
  \hat\varepsilon(E)  \equiv \frac{L_{\mathrm{eff}}^2}{s} \int\limits_{d}^{d+s} \frac{ \varepsilon(L,E)}{L^2} dL .
\label{equ:finaleff}
\end{equation}
If we choose $L_{\mathrm{eff}}=\sqrt{d (d+s)}$ as the geometric mean between the farest and nearest point baselines of the production straight, we have that $\hat\varepsilon(E)$ approaches unity if  $\varepsilon(L,E) \equiv 1$ or $L \gg s$ (far distance or point source approximation in \equ{ps}). The meaning of \equ{eavg} is the following: Implementing a near detector, we can use the same flux as for a conventional far detector, but with an effective baseline $L_{\mathrm{eff}} = \sqrt{d (d+s)}$, multiplied by the energy-dependent efficiency ratio $\hat\varepsilon(E)$ to be computed from the geometry of the source and detector. Therefore, for instance, for an implementation in the GLoBES software, only the effective baseline and  $\hat\varepsilon(E)$ is needed.

\begin{figure}[t]
 \begin{center}
 \includegraphics[width=0.31\textwidth]{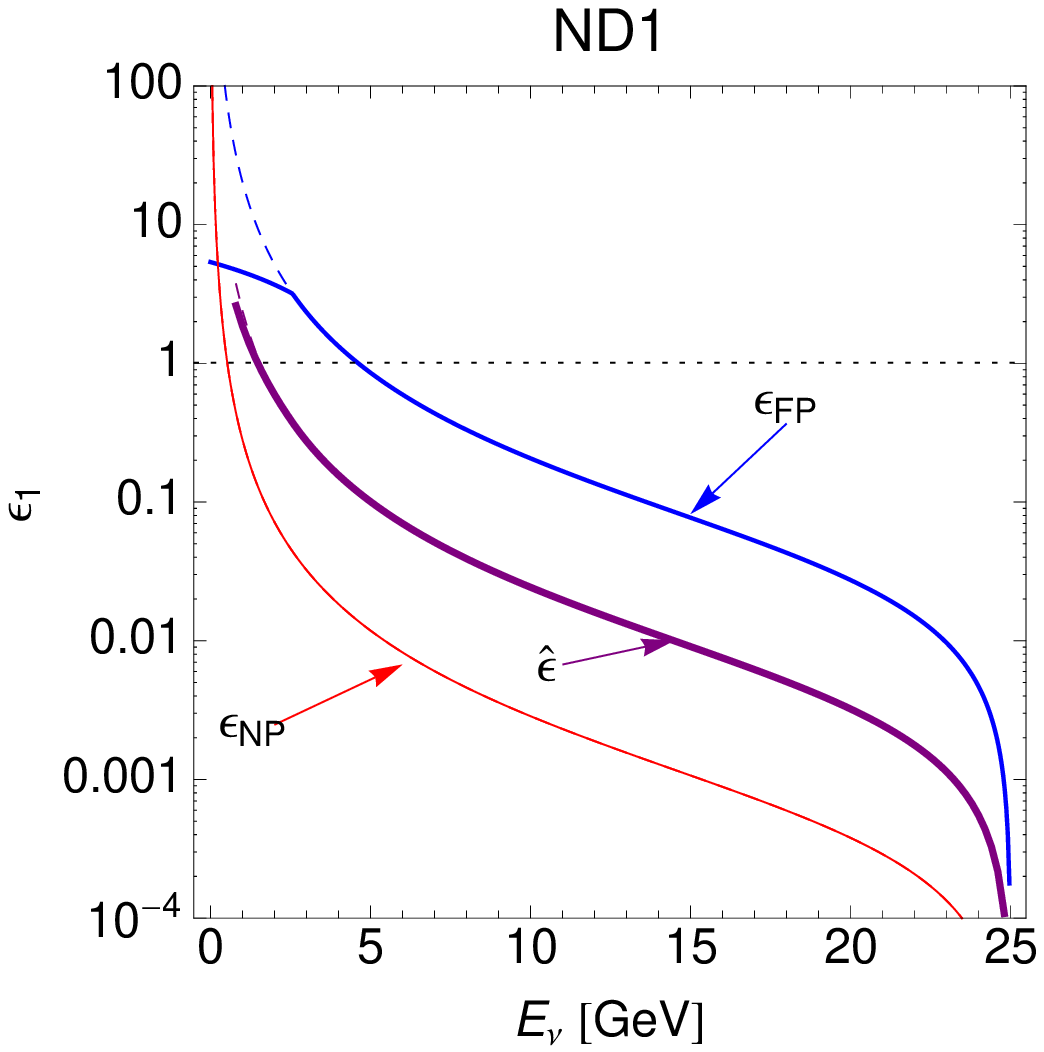} \hspace*{0.02\textwidth}
\includegraphics[width=0.31\textwidth]{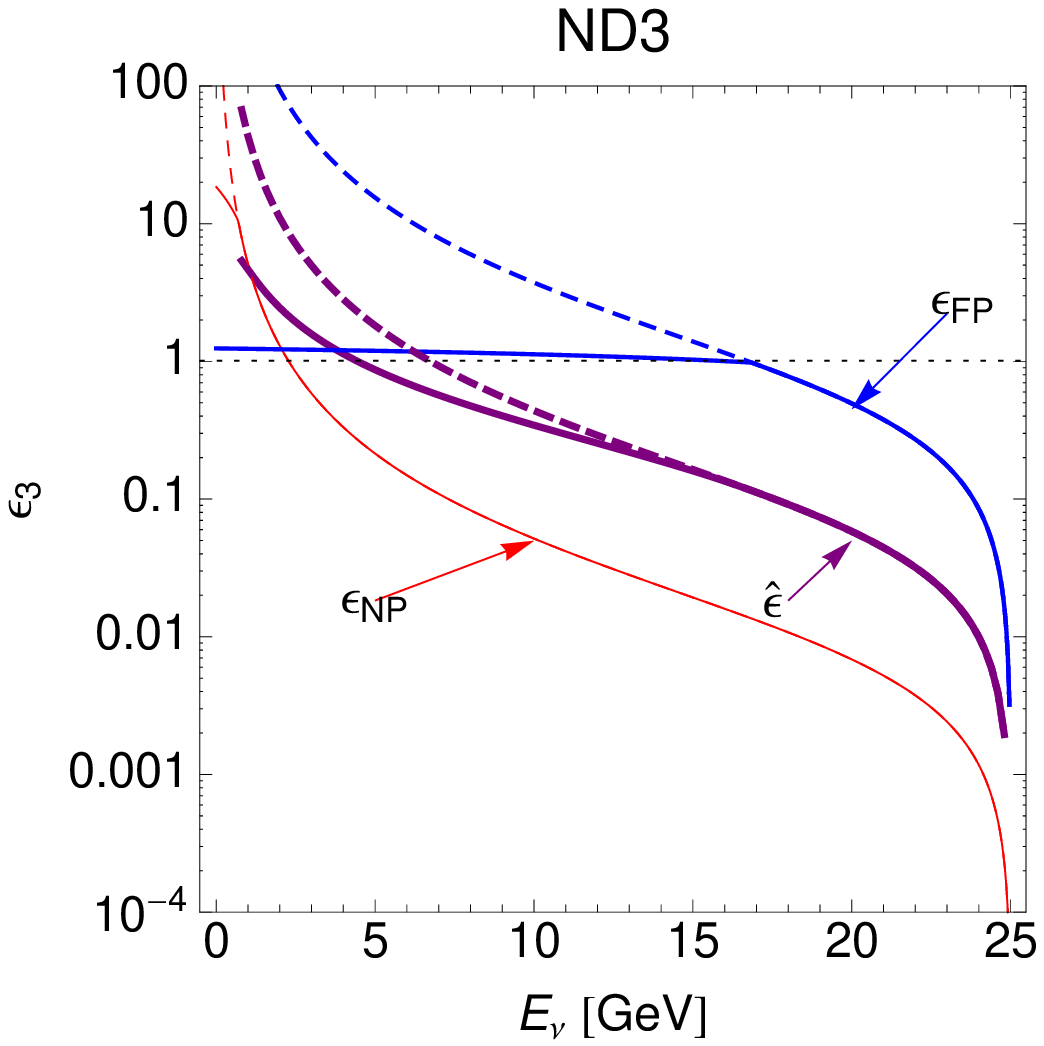}  \hspace*{0.02\textwidth}
\includegraphics[width=0.31\textwidth]{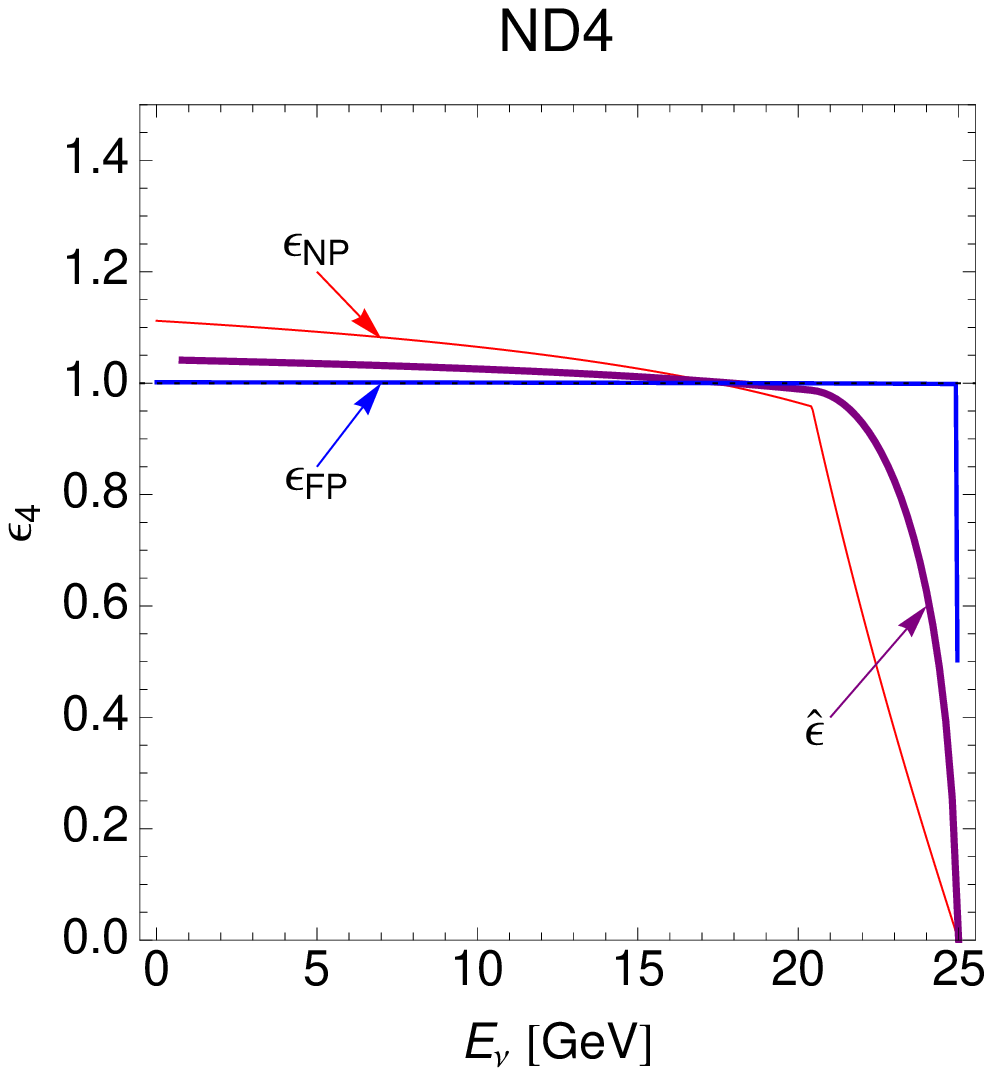}
 \end{center}
\mycaption{\label{fig:effs}The $\nu_\mu$ efficiency ratios as a function of $E_\nu$ for ND1, ND3, and ND4. The curves for the farest point ($\varepsilon_{\mathrm{FP}}$) and the nearest point ($\varepsilon_{\mathrm{FP}}$) to the near detector, as well as for the averaged efficiency $\hat\varepsilon$ are shown. The horizontal dotted lines express the far detector limit (same spectrum as on-axis spectrum), and the dashed curves the near detector limit (whole flux captured).}
\end{figure}

We show in \figu{effs} the efficiency ratio $\varepsilon(E)$ for the nearest point of the decay straight to the detector, the furthest point, and the average $\hat\varepsilon(E)$ for ND1, ND3, and ND4 and for $\nu_\mu$ (the efficiencies are flavor dependent). Obviously, the efficiency ratios decrease with energy, since for large energies the beam is in most cases smaller than the detector. For low energies, however, a part of the off-axis flux can be captured, leading to an increase of events compared to the on-axis case. The discontinuities in \figu{effs} come from \equ{thetacut}: If the beam, which has an energy dependent spread, becomes smaller than the detector, the efficiency ratio (or effective area) strongly decreases.
From \figu{effs}, we can read off that ND4 (and, similarly, ND2, which we do not show) performs similar to the far detector limit, since the efficiency ratios are close to one (even in the high energy part, at least within the energy resolution of the detector). ND1 has efficiency ratios strongly decreasing with energy, which means that the low energy part of the spectrum becomes enhanced. It corresponds to the near detector limit (whole beam captured, \cf, dashed curves), as anticipated. ND3 is an intermediate case: The farthest part of the decay straight resembles to the far detector limit, the nearest part the near detector limit.

\begin{figure}[t]
 \begin{center}
 \includegraphics[width=0.31\textwidth]{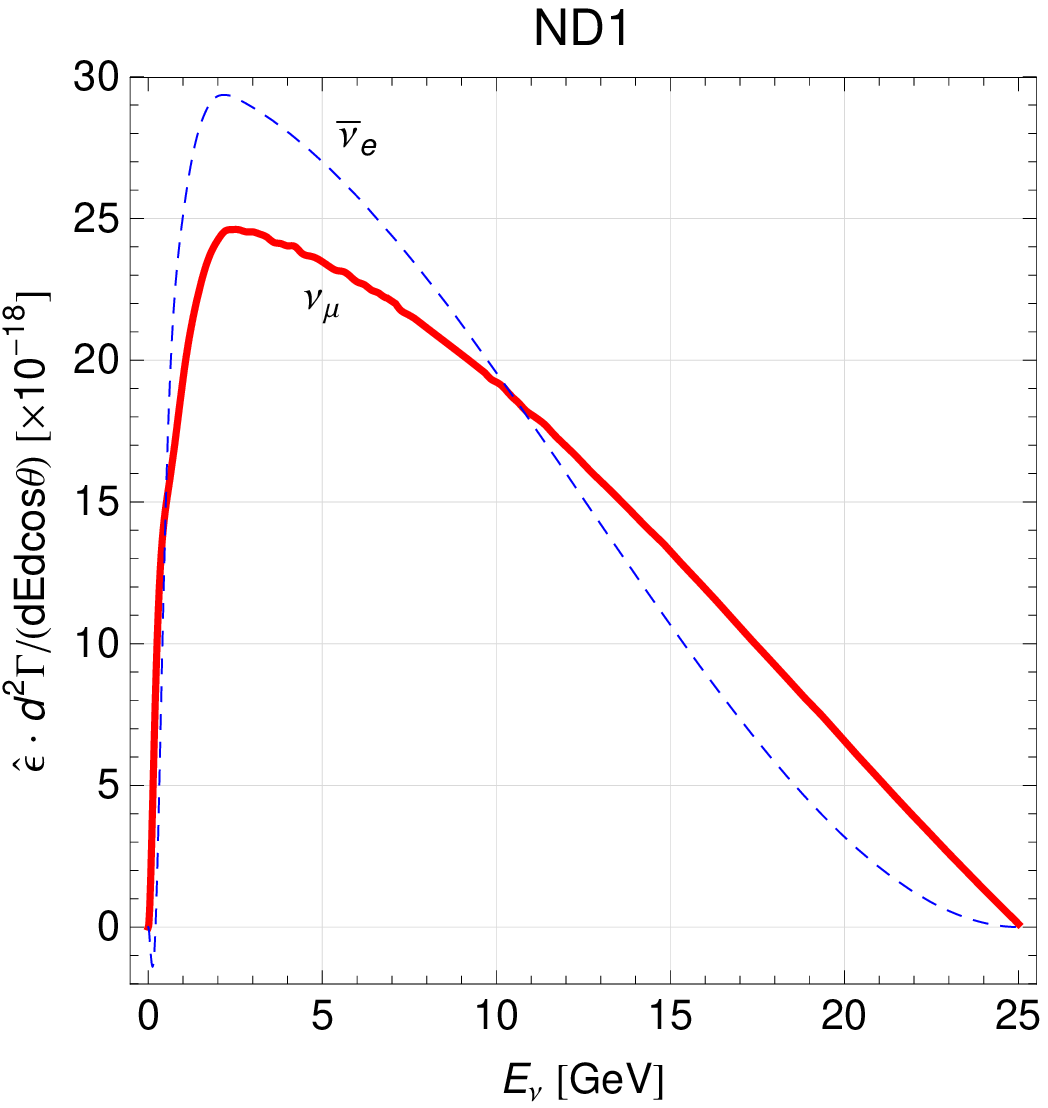} \hspace*{0.02\textwidth}
\includegraphics[width=0.31\textwidth]{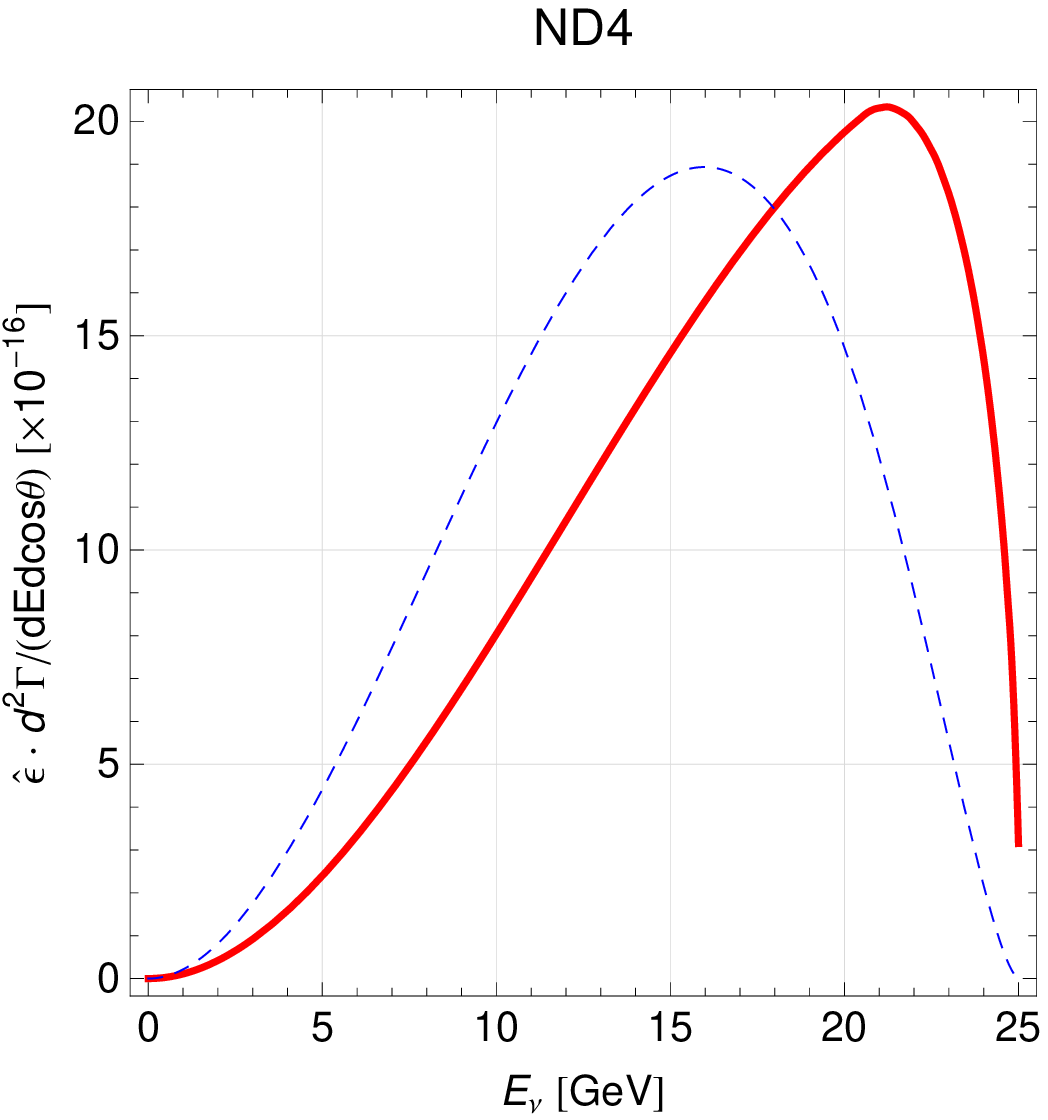} \hspace*{0.02\textwidth}
\includegraphics[width=0.31\textwidth]{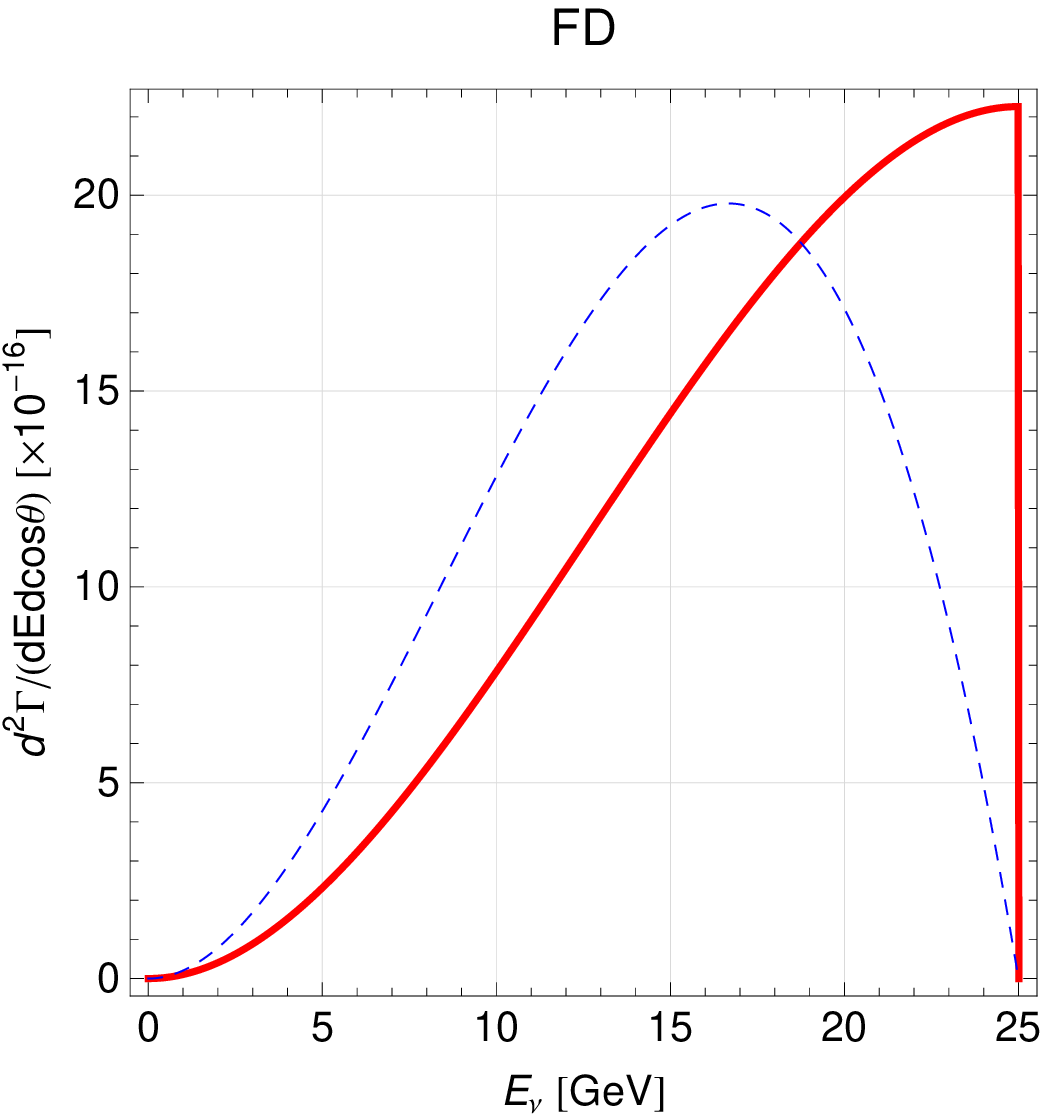}
 \end{center}
\mycaption{\label{fig:flux}The spectra (differential decay rates multiplied with $\hat\varepsilon$) for ND1, ND4, and the far detector (FD) limit, where the solid curves refer to the $\nu_\mu$ flux and the dashed curves to the $\bar\nu_e$ flux.}
\end{figure}

\begin{table}[t]
\begin{center}
\begin{tabular}{llr}
& Neutrinos ($\nu_\mu$) & Antineutrinos ($\bar\nu_\mu$) \\
\hline
ND1 & $1.71\times10^{10}$ & $8.82\times10^9$\\
ND2 & $3.69\times10^9$ & $1.93\times10^9$\\
ND3 & $1.39\times10^{10}$ & $7.21\times10^9$\\
ND4 & $9.36\times10^8$ & $4.89\times10^8$\\
\hline
\end{tabular}
\end{center}
\mycaption{\label{tab:events} Total event rates for ND1 to~4 for a high neutrino factory with $E_\mu=25 \, \mathrm{GeV}$ in ten years (see definition in \Sec~\ref{sec:syst}).}
\end{table}

In \figu{flux} the double differential decay rates similar to \figu{oaspectra} are shown, but multiplied with the corresponding efficiency factors $\hat\varepsilon$. Obviously, compared to the far detector case, the fluxes peak at lower energies, and the low energy part may be enhanced. Again, ND4 performs close to a far detector, ND1 peaks at lower energies. The other two near detectors are intermediate cases. The event rates for  ND1 to~4 are given in \Tab~\ref{tab:events}. Obviously, these rates are much higher than the ones in the far detector (a few hundred thousand in a distance $L=4000 \, \mathrm{km}$), which means that the detectors can in practice be built smaller (unless needed for other measurements, such as for particular cross section measurements).

\section{Systematics treatment and simulation methods}
\label{sec:syst}

Here we shortly describe the IDS-NF baseline setup including its systematics. Then we summarize our new systematics treatment, where details can be found in \App~\ref{app:syst}. Finally, we describe our simulation methods.

\subsection{IDS-NF baseline setup}

The currently discussed setup for a high energy neutrino factory is the IDS-NF baseline setup~\cite{ids}. It consists of two baselines $L_1 \simeq 4 \, 000 \, \mathrm{km}$
and $L_2 \simeq 7 \, 500 \, \mathrm{km}$, operated by two racetrack-shaped storage rings
simultaneously with both polarities ($\mu^+$ and $\mu^-$ stored, circulating
in different directions). The geometry of a storage ring is shown in \figu{ring}. There are no near detector specifications yet.
We define the polarities of the muons stored in the rings as:
\begin{eqnarray}
+ & : & \mu^- \rightarrow e^- + \bar\nu_e + \nu_\mu \\
- & : & \mu^+ \rightarrow e^+ + \nu_e + \bar\nu_\mu 
\end{eqnarray}
The following oscillation channels are used (with the corresponding muon polarities):
\begin{eqnarray}
\text{$\nu_\mu$ appearance }(-) & : & \nu_e \rightarrow \nu_\mu \label{equ:numuapp} \\
\text{$\bar\nu_\mu$ appearance }(+) & : & \bar\nu_e \rightarrow \bar\nu_\mu \label{equ:numubarapp} \\
\text{$\nu_\mu$ disappearance }(+) & : & \nu_\mu \rightarrow \nu_\mu \label{equ:numudisapp} \\
\text{$\bar\nu_\mu$ disappearance }(-) & : & \bar\nu_\mu \rightarrow \bar\nu_\mu \label{equ:numubardisapp}  
\end{eqnarray}
For the backgrounds, neutral currents are included for all channels, and mis-identified charged current events (from the disappearance channels) are included for the appearance channels. The background levels are about $10^{-4}$, depending on the energy (\cf, \Ref~\cite{ids} for details).
Since there are two racetrack-shaped storage rings $S1$ and $S2$ targeted towards two far detectors, there are altogether
eight oscillation channels.
In the IDS-NF baseline setup~1.0, the systematics treatment is rather straightforward.  For each channel and baseline, 
an overall normalization error is included, which is $2.5\%$ for the signal rates, and $20\%$ for the background rates. The normalization errors are uncorrelated between signal and background, among different channels, detectors, and polarities, but fully correlated among all bins. The energy resolution is assumed to be $\Delta E \, \mathrm{[GeV]} = 0.55 \sqrt{E \, \mathrm{[GeV]}}$.

\subsection{Refined systematics treatment}

We refine the IDS-NF systematics treatment by introducing the following systematical errors, focusing on the cross section measurement as main purpose of the near detectors:
\begin{description}
\item[Flux normalization] errors, fully uncorrelated among the different polarities $+$, $-$ and storage rings $S1$, $S2$,
but fully correlated among all bins and all channels operated with the same beam. For example, the flux normalizations in \equ{numuapp} and \equ{numubardisapp} are treated fully correlated, since they come from the same $\mu^+$ decay straight. Therefore, there are
four independent errors (two storage rings times two polarities). We assume that
the fluxes are known up to 2.5\% without knowledge from the near detectors, which is a very conservative starting hypothesis from the current IDS-NF systematics assumptions. However, we also test the impact of an improved flux knowledge of $0.1\%$, which may be obtained by using various beam monitoring devices~\cite{Abe:2007bi}.
\item[Cross section] errors for the inclusive charged current cross sections, fully correlated among all signal and background channels measuring $\nu_\mu$ or $\bar\nu_\mu$, but fully uncorrelated among all bins.\footnote{These errors are more accurately called ``shape errors'', because they are practically introduced after energy smearing (whereas cross section errors in principle enter before energy smearing). If the binning reflects the energy smearing, the cross section errors directly correspond to the shape errors.} We use
bin widths of 1~GeV from 1 to 10~GeV, of 2~GeV from 10 to 20~GeV, and of 2.5~GeV from 20 to 25~GeV, \ie, 16 bins in total. This means that we at least crudely follow the energy resolution of $\Delta E \, \mathrm{[GeV]} = 0.55 \sqrt{E \, \mathrm{[GeV]}}$.
Since we use 16 bins and $\nu_\mu$ and $\bar\nu_\mu$ cross sections, there are 32 uncorrelated errors. We adopt a conservative point of view and assumed that the cross section for each bin is externally known to about 30\% (see, \eg, summary plot in \Ref~\cite{Collaboration:2008pe}).
\item[Background normalization] errors, fully correlated among all bins, but fully uncorrelated among all channels, polarities, and detectors.\footnote{In principle, some knowledge on the backgrounds can be extrapolated from the near to the far detectors. However, the main backgrounds (charge identification and neutral current events) should rather depend on the charge identification and background discrimination properties of the far detector than the near-far extrapolation.
Therefore, we use this conservative approach.} In total, there are two (polarities) times two (baselines) times two (channels), making eight background errors in the far detectors. We assume a 20\% error each, just as in the IDS-NF baseline setup.
\end{description}
Compared to the IDS-NF systematics, the different signal errors are correlated in a particular way. For instance, the cross section errors at the two far detectors are fully correlated, which will turn out to have interesting effects.
For the efficiencies, however, we choose the IDS-NF numbers. Note that there might be some uncertainties coming from these efficiencies (such as a fiducial volume error), which we assume to be small and which we do not consider separately because they are not directly relevant for the near detector discussion. In addition, we neglect energy calibration errors.

For our near detectors ND1 to~4, we  introduce the same systematics for the backgrounds as for the far detectors, \ie, an uncorrelated 20\% error for each polarity. For the sake of simplicity, our near detectors have the same energy resolution and binning as the far detector.  However, as one important difference, we do not impose charge identification, since only the muon flavor is measured. Therefore, we assume 90\% MINOS-like efficiency, starting at a threshold of 1~GeV. Only NC backgrounds are considered, at the level of the far detectors. For details of the systematics treatment, see \App~\ref{app:syst}.

\subsection{Simulation techniques}

For the experiment simulation, we use the GLoBES software~\cite{Huber:2004ka,Huber:2007ji} with user-defined systematics. 
For the oscillation parameters, we use (see, \eg, \Refs~\cite{GonzalezGarcia:2007ib,Schwetz:2008er}) $\sin^2 \theta_{12}=0.3$, $\sin^2 \theta_{23} = 0.5$, $\sdm = 8.0 \cdot 10^{-5} \, \mathrm{eV^2}$, $\ldm = 2.5 \cdot 10^{-3} \, \mathrm{eV^2}$, and a normal mass hierarchy, unless specified otherwise. We impose external errors on $\sdm$ and $\theta_{12}$ of 4\% each, and we include a 2\% matter density uncertainty~\cite{Geller:2001ix,Ohlsson:2003ip}.

The fluxes of the near detectors are included in GLoBES as user-defined fluxes. The user-defined fluxes are obtained from the product between the on-axis (normalized) rate \equ{nps} and the energy-dependent efficiencies in \equ{finaleff}.  Note that for the baseline, $L_{\mathrm{eff}}$ has to be used.
The event rate in GLoBES, similar to the one described in \equ{nps}, is scaled automatically with the factors $n_\mu$, $\sigma$, and $M_{\mathrm{Det}}$.\footnote{For the sake of simplicity, we have not included time in the discussion, but only discuss the time-integrated event rates.} In order to obtain the rest of the normalization in proper units, the normalization factor is determined as in App.~C of the GLoBES manual in excellent agreement with the built-in fluxes:
\begin{align}
 @norm&=\frac{1}{2\pi \times (1 \, [\text{km}])^2} \times \frac{1}{10^{10} \, [{\rm cm}^{2}/{\rm km}^{2}]}\times 1 \, [{\rm kton}] \times 10^{9} \, [{\rm g/kton}]\times \frac{1}{m_N [{\rm g}]} \times 10^{-38} \, [{\rm cm}^2]\nonumber\\
&=0.9581\times10^{-16} \, .
\end{align}
Here $m_N$ is the average nucleon mass of the target material, \ie, $1/(m_N [{\rm g}])$ is the number of target nuclei in one gram of target material. The first factor comes from the factor $1/(2 \pi L^2)$ in \equ{nps}, \ie, the conversion from flux per angle to flux per surface area (the flux in GLoBES has to be specified in a distance of 1~km as flux per surface area).

\section{Standard oscillation physics with a high energy neutrino factory}
\label{sec:so}

In this section, we perform a qualitative discussion of the impact of our systematics
on different measurements, and we show several quantitative examples.

\subsection{Qualitative discussion}
\label{sec:qd}

\begin{figure}[t]
\includegraphics[width=\textwidth]{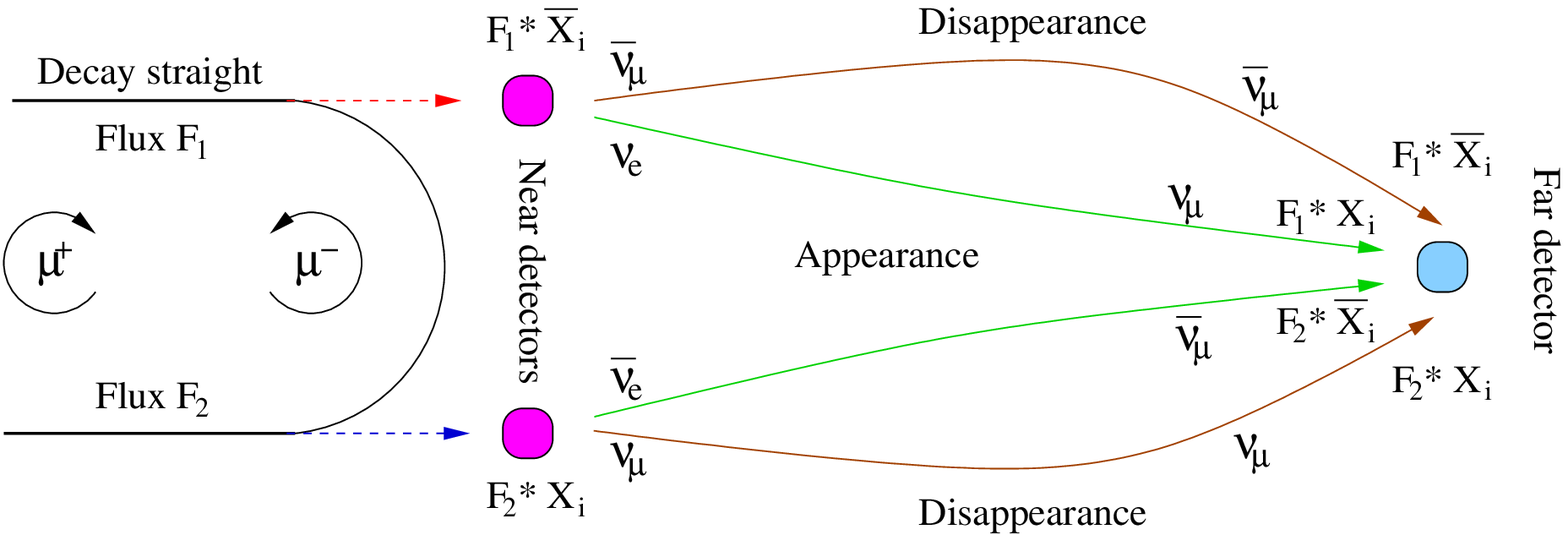}
\mycaption{\label{fig:ringch} Simplified systematics for one storage ring. Here $X_i$ refers to the $\nu_\mu$ cross section, $\bar X_i$ to the $\bar\nu_\mu$ cross section, and $F_1$ and $F_2$ to the flux normalizations. The
index $i$ marks the bin-dependence of the cross sections, whereas the flux normalizations are assumed to be bin-independent.}
\end{figure}

For the following discussion, it is useful to have a simplified picture of systematics, which we illustrate  in \figu{ringch} for one storage ring, \ie, one baseline. Here $X_i$ refers to the $\nu_\mu$ cross section, $\bar X_i$ to the $\bar\nu_\mu$ cross section, and $F_1$ and $F_2$ to the flux normalizations for $\mu^+$ and $\mu^-$ stored, respectively. The
index $i$ illustrates the bin-dependence of the cross sections, whereas the flux normalizations are assumed to be bin-independent. The figure shows what products of cross sections and fluxes are measured in the different detectors.

From the physics point of view, one may distinguish three different classes of measurements, which we have also tested quantitatively:

{\bf  Measurement of the leading atmospheric parameters.}
The main sensitivity for the leading atmospheric parameters $\ldm$ and $\theta_{23}$ comes from
the disappearance channel. From \figu{ringch}, we read off that only the product $F_1 \, \bar X_i$ (or $F_2 \, X_i$ for the different polarity) is measured in both the near and far detectors. Therefore, it is not necessary to disentangle $F_1$ from $\bar X_i$, and the product is precisely determined by the near detectors in each energy bin. For a large enough near detectors, this measurement is hence limited by the statistics of the far detector. However, without near detectors, the shape error coming from the cross section uncertainties will highly affect the shape-sensitive disappearance measurement.

{\bf Measurements at the sensitivity limit of the appearance channel.}
At the sensitivity limit of the appearance channel, such as for the $\theta_{13}$ and mass hierarchy measurements, the measurement will be limited by the backgrounds and their uncertainties. Therefore, even large uncertainties in the cross sections or fluxes do not harm the measurements, and the near detectors will not be required within our systematics treatment (which we have explicitely tested). However, it could help to better understand the backgrounds, a subject, which is beyond the scope of this study.

{\bf CP violation measurement with the appearance channel.}
In this case, one has to distinguish the (moderately) small $\theta_{13}$ and large $\theta_{13}$ limits. In the small $\theta_{13}$ case, the measurement is background dominated, where ``background'' refers to both systematics and the CP-invariant contribution to the appearance probability (see, \eg, Eq.~(31) in \Ref~\cite{Akhmedov:2004ny}). This contribution depends, for instance, on $\theta_{23}$. Since the leading atmospheric parameters cannot be determined precisely without near detectors (see above), there will be some use of the near detectors no matter if the flux is known or not, \ie, the cross section uncertainties will be translated into a background uncertainty.

For large enough statistics in the appearance channel, the knowledge on the signal normalization (such as on cross sections and fluxes) becomes an important factor. 
In this case, the products $F_1 \, \bar X_i$ and $F_2 \, X_i$ will be determined by the near detectors (\cf, \figu{ringch}). However, at the far detector, the different products $F_1 \, X_i$ and $F_2 \, \bar X_i$ are needed to extract the information. Therefore, the ratio between the neutrino and antineutrino events, which is directly related to the CP violation effect, can be
written as
\begin{equation}
\underbrace{\frac{N_\nu}{N_{\bar\nu}}}_{\mathrm{FD \, measured}} \propto \frac{(F_1 \, X_i)}{(F_2 \, \bar X_i)} \, \frac{P(\nu_e \rightarrow \nu_\mu)}{P(\bar\nu_e \rightarrow \bar\nu_\mu)} = \underbrace{\frac{(F_1 \, \bar X_i)}{(F_2 \, X_i)}}_{\mathrm{ND \, \, measured}} \, \frac{X_i^2}{\bar X_i^2} \,  \frac{P(\nu_e \rightarrow \nu_\mu)}{P(\bar\nu_e \rightarrow \bar\nu_\mu)}  \, 
\end{equation}
where the first factor in the last product is measured by the near detectors. This means that only the product 
$X_i^2/\bar X_i^2 \times  P(\nu_e \rightarrow \nu_\mu)/P(\bar\nu_e \rightarrow \bar\nu_\mu)$ can be extracted, which leads to a different CP violation interpretation depending on the cross sections. In this case, there is an  anti-correlation between fluxes and cross sections determined by $F_1 \, \bar X_i =C_1=const.$ and $F_2 \, X_i =C_2=const.$ in the near detectors independent of the energy bin $i$, and the cross sections can only be extracted from these relationships to the degree the fluxes are known. 
In the following, we assume that the fluxes can be monitored at the level of 0.1\% {\bf if the near detectors are present}, \ie, this anti-correlation can be broken. For instance, the event rates from purely leptonic elastic scattering interactions, for which the cross sections can be exactly calculated, should be sufficiently high to provide this precision~\cite{Abe:2007bi}.

\subsection{Measurement of the atmospheric parameters}

\begin{figure}[t]
\begin{center}
\includegraphics[width=\textwidth]{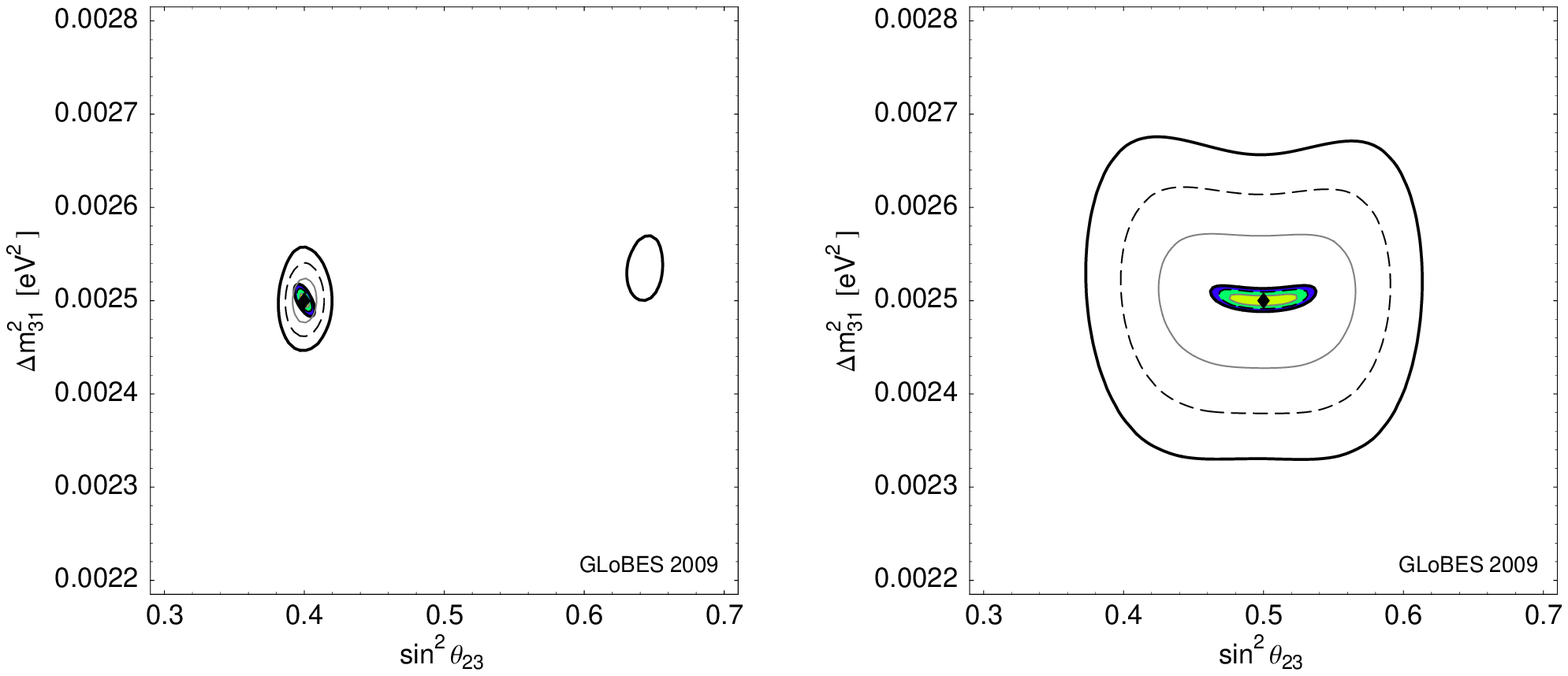}
\end{center}
\mycaption{\label{fig:atmhigh} The $\sin^2 \theta_{23}$-$\ldm$ allowed region  for a high energy neutrino factory at the $L=4000 \, \mathrm{km}$ baseline only; $1\sigma$, $2 \sigma$, $3\sigma$ CL (2 d.o.f.), best-fit points marked by diamonds. The filled contours correspond to our near detector-far detector simulation, whereas the unfilled contours represent the far detector only.
Normal hierarchy only, \ie, no sign-degenerate solution shown. In the left panel, $\stheta=0.08$ and $\deltacp=0$, in the right panel, $\stheta=0$.}
\end{figure}

As already discussed above, the near detectors are very important for the atmospheric parameter measurements, which are very sensitive to spectral effects. The impact of the near detectors is illustrated in \figu{atmhigh} for a single baseline neutrino factory ($L=4000 \, \mathrm{km}$). The unfilled contours represent the far detector only, whereas the filled contours represent the far-near combination. In the right panel, the effect of the near detectors is very large for maximal mixing. Note that the unfilled contours depend on the knowledge of the cross sections, whereas the filled contours are limited by the statistics in the far detector. In the left panel, a non-maximal value of $\sin^2 \theta_{23}$ is chosen. In this case, the effect is less dramatic than in the right panel for maximal mixing, but still substantial. Especially, the octant degeneracy can be excluded with the near detectors at a high confidence level (if $\stheta$ is large enough).

\begin{figure}[t]
\begin{center}
\includegraphics[width=\textwidth]{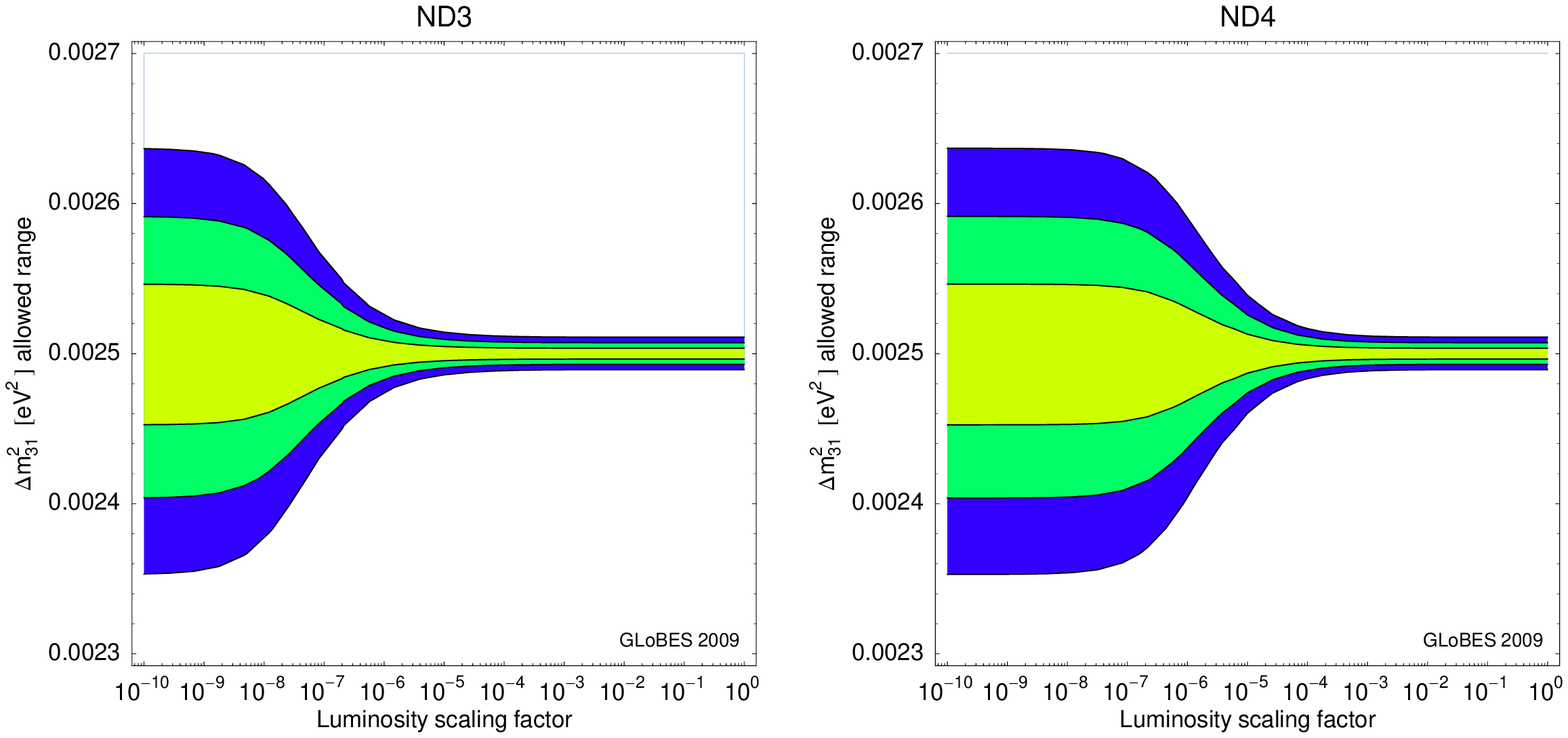}
\end{center}
\mycaption{\label{fig:lumiscale} The allowed range for $\ldm$ as a function of a luminosity scaling factor rescaling the near detector masses $\times$ operation time (for ND3 and ND4). The contours are shown for the $1\sigma$, $2 \sigma$, $3\sigma$ CL (1 d.o.f.). For $L_1=4000 \, \mathrm{km}$ and normal hierarchy only (\ie, no sign-degenerate solution shown), for maximal mixing and $\stheta=0$.}
\end{figure}

The results hardly depend on which of ND1 to ND4 is chosen, because they all have sufficient statistics in all bins. This is illustrated in \figu{lumiscale}, where the $\ldm$ allowed range is shown as a function of a luminosity scaling factor rescaling the near detector masses $\times$ operation time for two of the near detectors (ND3 and ND4). Comparing the large number of events in our near detectors  in \Tab~\ref{tab:events} with the few hundred thousand events in the far detector (disappearance channel), it is not surprising that even much smaller near detectors would do. Basically, the luminosity starts to become important at the point when the near detector rates are of the order of the far detector rate. From \figu{lumiscale} and \Tab~\ref{tab:nd}, we can read off at the example of ND4 that, in principle, near detectors with a fiducial mass below 1~kg are sufficient for the standard oscillation parameter measurements. In this case, the near detectors can be easily operated in the far distance limit, \ie, the first part of \equ{ps} applies.
Because there is no significant difference among the different near detectors, we only refer to the ``near detectors'' in the following. 

Very interestingly, if the neutrino factory is operated with two baselines at $L_1=4000 \, \mathrm{km}$ and $L_2=7500 \, \mathrm{km}$, there is no significant effect of the near detectors -- despite the fact, that two additional flux normalization errors for the second storage ring are introduced. Obviously, the spectral errors can be resolved by measuring the same cross sections in two different far detectors similar to a near-far combination, and energy-independent normalization errors are of secondary importance.

\subsection{CP violation measurement}

\begin{figure}[t]
\begin{center}
\includegraphics[width=\textwidth]{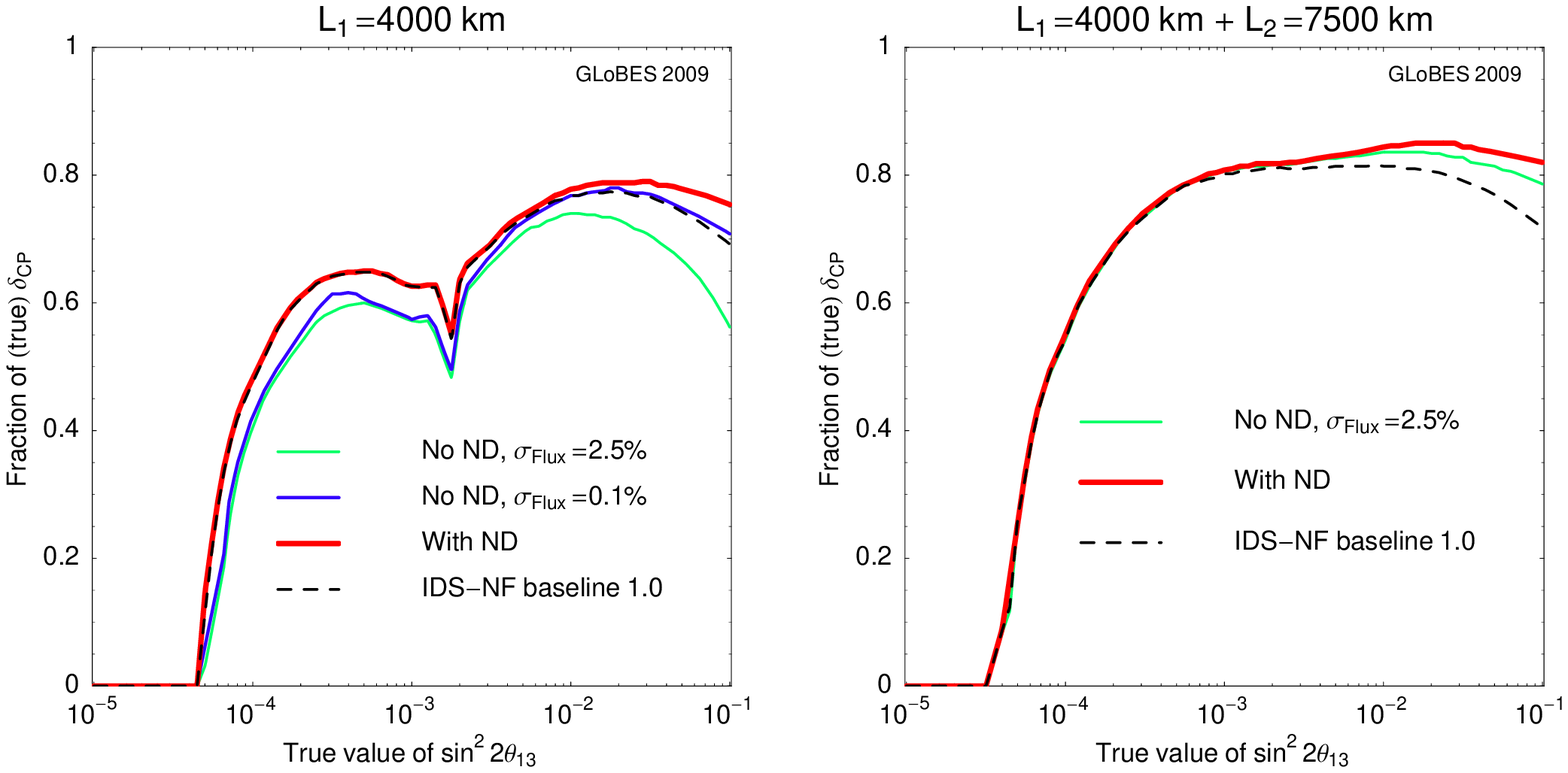}
\end{center}
\mycaption{\label{fig:ndcomp} CP violation discovery reach as a function of true $\stheta$ and the fraction of (true) $\deltacp$ for one far detector (left) and two far detectors (right); $3\sigma$ CL.}
\end{figure}

We show in \figu{ndcomp} the CP violation discovery reach as a function of true $\stheta$ and the fraction of (true) $\deltacp$ for one far detector (left) and two far detectors (right). The dashed curves represent the IDS-NF baseline setup with the corresponding systematics treatment, the solid lower curves our systematics treatment without near detectors and flux uncertainties $\sigma_{\mathrm{Flux}}=2.5\%$, and the solid (thick) upper curves our systematics treatment with near detectors. In the left plot, we include a curve with a better known flux, whereas in the right plot, this curve coincides with the thick curve.
As first important observation, our systematics treatment (including near detectors) leads to a better sensitivity that the IDS-NF systematics, in spite of our conservative choices for the systematical uncertainties. The main difference between these two systematics treatments is that we assume the systematical errors to be correlated among different detectors, whereas the IDS-NF setup assumes fully uncorrelated errors. The practical realization will likely be somewhere in the middle (between the dashed and thick curves), since, for instance, cross sections are fully correlated among all detectors measuring the same flavor and polarity, whereas there may be other normalization and spectral errors which depend on the detector (such as fiducial volume errors leading to errors in the detector-dependent efficiencies).
Similar to the atmospheric parameter measurement, the near detectors are important for one baseline only, whereas a two-baseline neutrino factory hardly benefits from the near detectors. 

For one baseline (left plot), an improved flux knowledge only helps for large $\theta_{13}$ in the absence of the near detectors, because the correlation described in \Sec~\ref{sec:qd} can be resolved and only the uncertainty of the atmospheric parameter measurements remains.
We have also tested a better matter density knowledge of 0.5\% (compared to 2\%), which slightly improves the measurement in all cases for large $\theta_{13}$. In the case of two baselines with near detectors in the shorter baseline storage ring (thick curve in right plot), we have found no improvement from better known fluxes. Since the cross sections can in this case already be extracted by the two near detectors in the first storage ring, this implies that putting additional near detectors in front of the second storage ring, such as for flux monitoring, may not be required from the physics point of view. From the figure, neither a precise flux monitoring nor near detectors are mandatory for a successful CP violation measurement in a two baseline neutrino factory, if the far detectors are well enough understood.

\section{Standard oscillation physics with a low energy neutrino factory}
\label{sec:solow}

Here we briefly discuss the use of near detectors for a low energy neutrino factory.
For the low energy neutrino factory, there is not yet any baseline specification
within the IDS-NF. Therefore, for our reference setup, we follow \Ref~\cite{Bross:2007ts},
using $E_\mu=4.12 \, \mathrm{GeV}$ and their ``high statistics'' scenario with $10^{23} \, \mathrm{kt} \, \mathrm{yr}$
exposure, corresponding to 10 years of data taking with $5 \cdot 10^{20}$ useful muon decays per polarity and year,
and a fiducial detector mass times efficiency of $20 \, \mathrm{kt}$. We use the baseline $L=1290 \, \mathrm{km}$, which corresponds to Fermilab-Homestake. The detector proposed in this reference
is a magnetized TASD (Totally Active Scintillating Detector) with a threshold of about $500 \, \mathrm{MeV}$,
an energy resolution $\Delta E \, \mathrm{[GeV]} = 0.3 \, E \, \mathrm{[GeV]}$ (conservative estimate), and
an estimated detection efficiency of 73\%. For the binning, we use nine equi-distant bins between  $500 \, \mathrm{MeV}$
and $5 \, \mathrm{GeV}$. The channels are the same as used for the high energy version, but the background level is conservatively estimated to be $10^{-3}$~\cite{Bross:2007ts}. Note that we include two types of backgrounds for the appearance channels, one which scales with the disappearance rates (such as from charge mis-identification), and one which scales with the un-oscillated spectrum (such as from neutral current events), both at the level of $10^{-3}$.
As in  \Ref~\cite{Bross:2007ts}, we take the systematical errors to be 2\% for all signal and background errors
in the reference setup. 
We can reproduce the event rates in \Ref~\cite{Bross:2007ts} very well, as well as the fits in the $\theta_{23}$-$\ldm$-planes and $\theta_{13}$-$\deltacp$-planes for their set of parameters. Compared to \Ref~\cite{Bross:2007ts}, we do not find any sensitivity to the mass hierarchy for $\theta_{13}=0$.

For the near detector implementation, we follow the same procedure as for the high energy neutrino factory using ND3.
However, we only have one storage ring and one far detector, and nine (equi-distant) bins, making a total of 18 cross section errors (instead of 32). The external knowledge of the fluxes $\sigma_{\mathrm{Flux}}$ is assumed to be 2\% (if not stated otherwise), and the backgrounds are assumed to be known at the same level~\cite{Bross:2007ts}. 

\begin{figure}[t]
\begin{center}
\includegraphics[width=\textwidth]{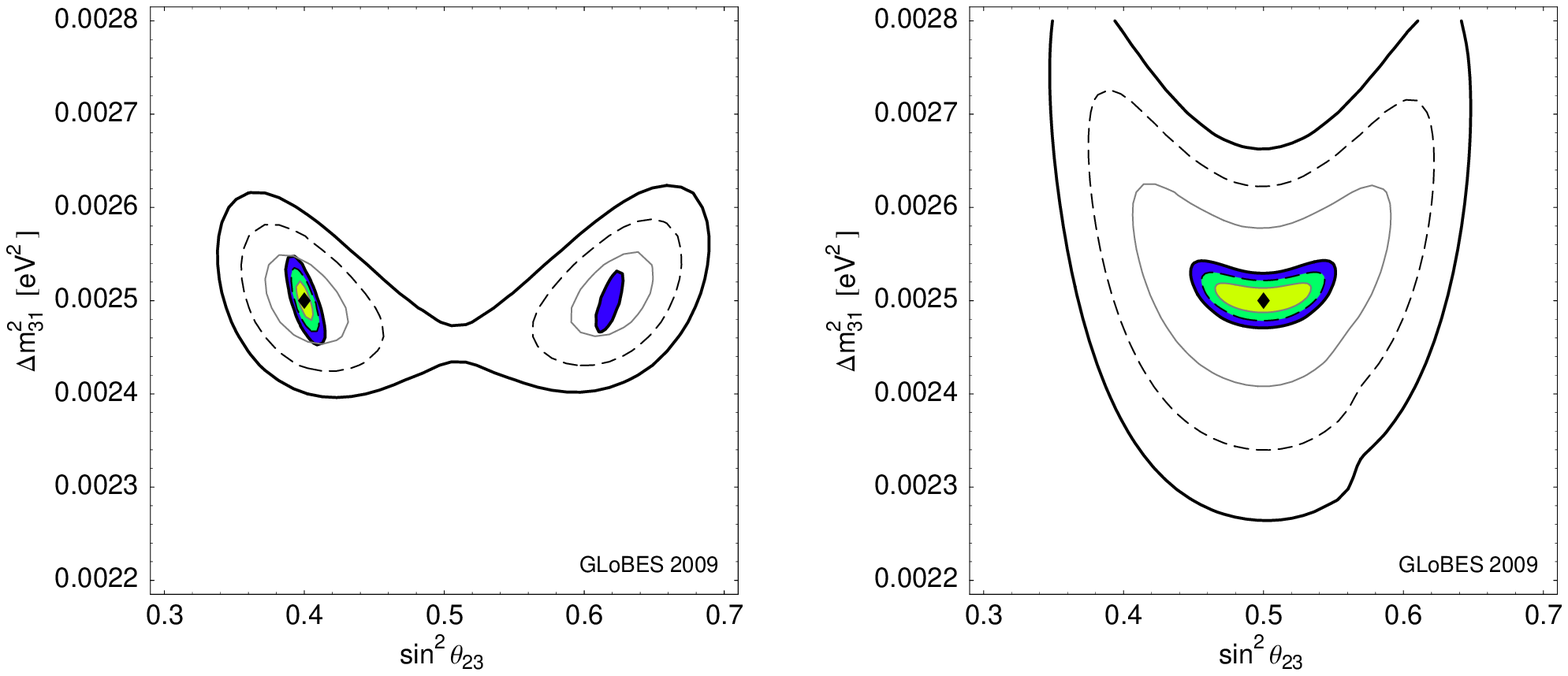}
\end{center}
\mycaption{\label{fig:atmlow} The $\sin^2 \theta_{23}$-$\ldm$ allowed region  for a low energy neutrino factory; $1\sigma$, $2 \sigma$, $3\sigma$ CL (2 d.o.f.), best-fit points marked by diamonds. The filled contours correspond to our near detector-far detector simulation, whereas the unfilled contours represent the far detector only.
Normal hierarchy only, \ie, no sign-degenerate solution shown. In the left panel, $\stheta=0.08$ and $\deltacp=0$, in the right panel, $\stheta=0$.}
\end{figure}

In \figu{atmlow}, we show the results for the leading atmospheric parameters for a deviation from maximal mixing and large $\theta_{13}$ (left), and maximal mixing (right). The unfilled contours correspond to no near detector, and the filled contours include the near detectors. As for the high energy neutrino factory with one baseline only, the near detectors are very important for the atmospheric parameter measurement. Of course, the results without near detectors will depend on the assumptions on the cross sections. However, one can easily see the benefits of the near detectors, especially since there is typically no second baseline planned for the low energy neutrino factory.

\begin{figure}[t]
\begin{center}
\includegraphics[width=0.5\textwidth]{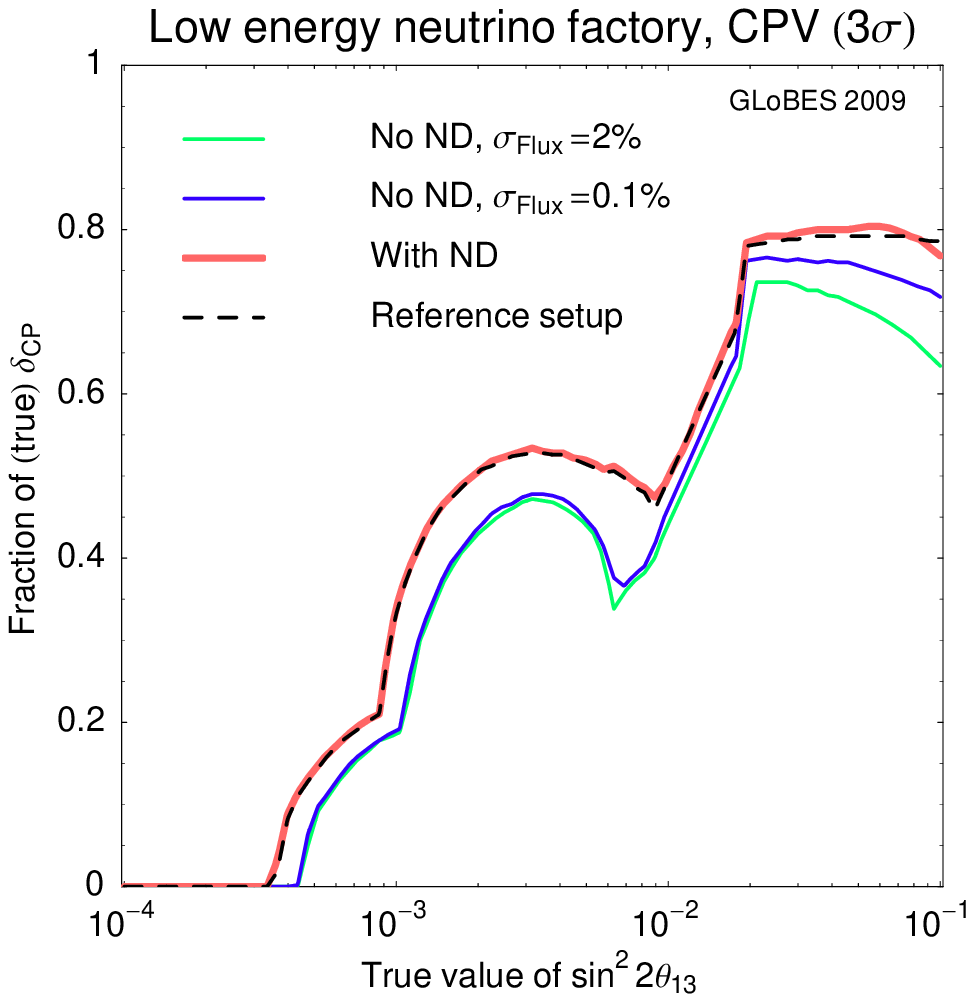}
\end{center}
\mycaption{\label{fig:lowenufact} CP violation discovery reach as a function of true $\stheta$ and the fraction of (true) $\deltacp$ for a low energy neutrino factory; $3\sigma$ CL. The reference setup is taken from \Ref~\cite{Bross:2007ts}.}
\end{figure}

In \figu{lowenufact}, we show the main results for CPV (where the effect is most dramatic). Obviously, the setup with near detectors corresponds very much to the reference setup~\cite{Bross:2007ts}, except for large $\theta_{13}$, where the reference setup is slightly better. Without near detectors, the result is significantly worse, which means that the near detectors are mandatory for a low energy neutrino factory of all $\theta_{13}$ to better understand the cross sections. In either case (with or without near detector), a better known flux (0.1\%) helps somewhat for large $\stheta$, whereas the matter density uncertainty is of secondary importance for this setup. As for the high energy neutrino factory with only one far detector, a better known flux yields almost no improvement for small $\theta_{13}$.

\section{Non-standard interactions with a high energy neutrino factory}
\label{sec:nsi}

Non-standard interactions (NSI)~\cite{Wolfenstein:1977ue,Valle:1987gv,Guzzo:1991hi,Grossman:1995wx,Roulet:1991sm}  are effects of physics beyond the Standard Model
(SM) which can be described by four fermion vertices below the electroweak symmetry breaking scale
relative to the SM interactions:
\begin{equation}
\delta \mathscr{L}_{\text{eff}}
= 2 \, \sqrt{2} \, G_F \, (\epsilon^{L/R})^{\alpha \gamma}_{\beta \delta} \, 
 \left(
 \bar{\nu}^{\beta} \gamma^{\rho} {\rm P}_{L} \nu_{\alpha}
  \right) \, 
 \left(
 \bar{\ell}^{\delta} \gamma^{\rho} {\rm P}_{L/R} \ell_{\gamma} 
 \right) \, .
\label{equ:nsi}
\end{equation}
Here $P_L$ and $P_R$ are the the left- and right-handed 
(chiral) projection operators, respectively. The operator in \equ{nsi} 
may come from effective non-renormalizable $d=6$~\cite{Buchmuller:1985jz,Bergmann:1998ft,Bergmann:1999pk}, $d=8$~\cite{Berezhiani:2001rs,Davidson:2003ha},
 \etc, operators 
using the SM field content (including the Higgs). The
$\epsilon$ from the dimension $d$ operator is suppressed by $\epsilon \propto (v/\sqrt{2}\Lambda)^{d-4}$
with respect to the SM Higgs vacuum expectation value $v$ by the new physics scale $\Lambda$, which comes from
integrating out the heavier (new) fields. 

These NSI can affect the neutrino propagation by coherent forward
scattering in Earth matter if the two charged leptons
in \equ{nsi} are electrons,  which is only sensitive to the vector component as
\begin{equation}
\epsilon^{m}_{\beta \alpha} = \epsilon^{m,L}_{\beta \alpha} +  \epsilon^{m,R}_{\beta \alpha} \quad \mathrm{with}
 \quad \epsilon^{m,L/R}_{\beta \alpha} = (\epsilon^{L/R})^{\alpha e}_{\beta e} \, .
\end{equation}

In addition to the propagation in matter, the production or detection processes can be affected by NSI. For the specific case of a neutrino factory and considering just purely leptonic NSI, dominantly effects at the source are relevant, since the (initial) detection interactions involve quarks. They are customarily parameterized in terms of $\epsilon_{\alpha \beta}^s$, which describes an effective source state $ |\nu^s_\alpha \rangle$ (for small $\epsilon^s$) as~\cite{Grossman:1995wx, Gonzalez-Garcia:2001mp, Bilenky:1992wv} 
\be \label{equ:nsisource}
|\nu^s_\alpha \rangle = |\nu_\alpha\rangle + \sum_{\beta=e,\mu,\tau}\epsilon^s_{\alpha\beta} |\nu_\beta\rangle \, 
 \quad \mathrm{with} \quad \epsilon^{s}_{\mu \beta} 
 =
(\epsilon^L)^{e \mu}_{\beta e} \quad \text{or} \quad
\epsilon^{s}_{e \beta} 
 =
(\epsilon^L)^{\mu e}_{\beta \mu}\,.
\ee 
 In this case, the muon decay rate could be modified by the NSI interaction in \equ{nsi}, with the largest effect resulting from the coherent contribution to the state at the source~\cite{Gonzalez-Garcia:2001mp, Huber:2002bi}.
It appears as an admixture of a given flavor $\nu_\alpha$ with all other flavors, encoded by $\nu_\beta$ in \equ{nsisource}. The dominating effect comes from the left-handed component, since the right-handed component will be helicity suppressed at production.

Near detectors may be very relevant for the extraction of the source NSI, since these lead to an unambiguous ``zero-distance'' ($L \ll L^{\mathrm{osc}}$) effect proportional to $|\epsilon^s|^2$. However, within our systematics framework, how should one disentangle source NSI modifying the electron or muon flavor contributions ($\epsilon^s_{e e}$, $\epsilon^s_{e \mu}$, $\epsilon^s_{\mu e}$, and $\epsilon^s_{\mu \mu}$)  from the unknown fluxes and cross sections? For example, a larger than anticipated muon neutrino rate at zero distance may come from underestimated cross sections, the neutrino flux normalization, or $\epsilon^s_{\mu \mu}$, \ie, it will be very difficult to disentangle these contributions, especially if more of the $\epsilon$'s are allowed simultaneously. For $\epsilon^s_{e \mu}$, however,  near detectors with charge identification could help (because there will be $\nu_\mu$ of the opposite polarity in the beam), and for $\epsilon^s_{\mu e}$ near detectors with electron flavor measurement and charge identification. Since our near detectors are more primitive than that, we do not consider these measurements. However, note that for some new physics measurements, such as electron neutrino disappearance into sterile neutrinos~\cite{Giunti:2009zz}, these additional channels are important.

\begin{figure}[t]
\begin{center}
\includegraphics[width=0.5\textwidth]{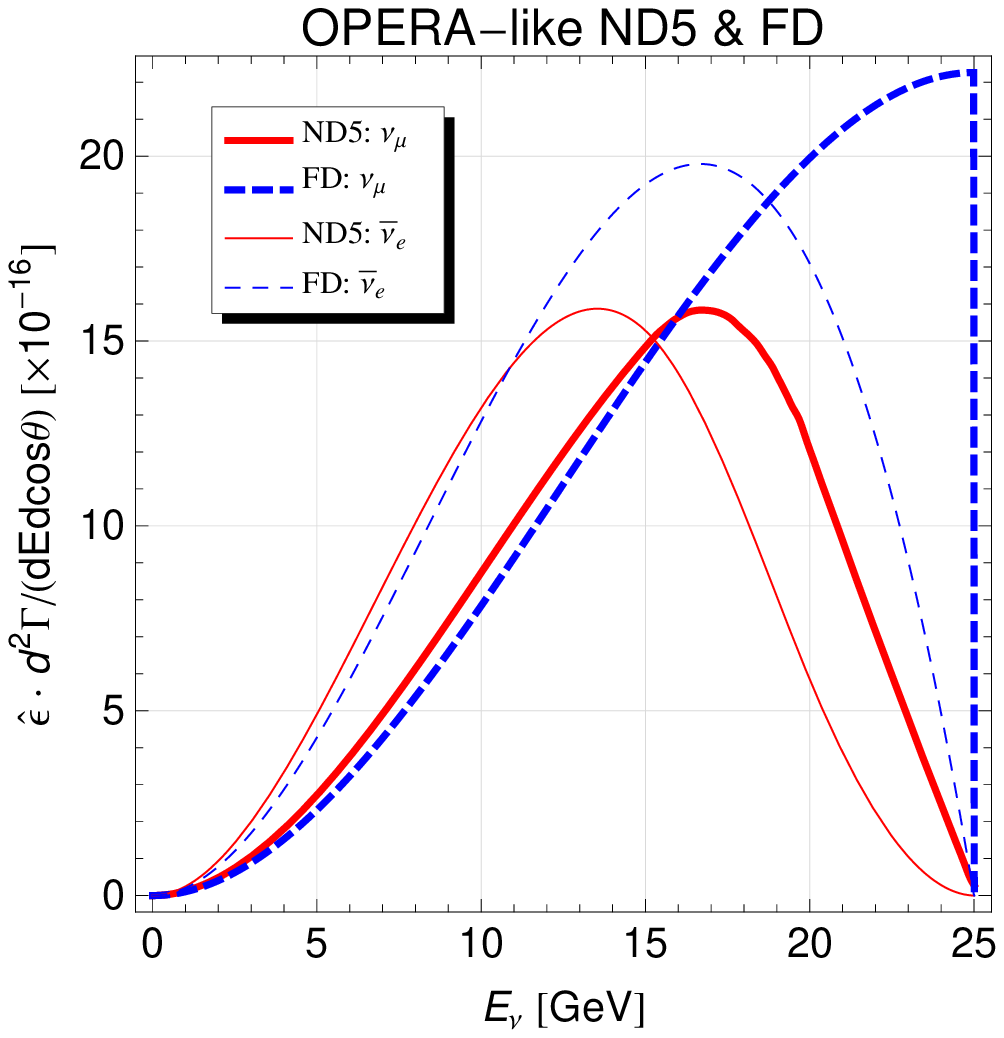}
\end{center}
\mycaption{\label{fig:flux5} Spectra (differential decay rates multiplied by the efficiency factor $\hat\varepsilon$) for the $\nu_\mu$ and $\nu_e$ fluxes of the OPERA-sized ND5 ($d=1 \, \mathrm{km}$). For comparison, the spectra of the far distance (FD) limit are shown.}
\end{figure}

Because there are no tau neutrinos in the beam, the most interesting option may be to use near detectors to measure $\nu_\tau$ appearance, see, \eg, \Refs~\cite{Malinsky:2008qn,Malinsky:2009gw}.  Therefore, we use the high energy neutrino factory with ND3 and two (symmetrically operated) additional near detectors at $L=1 \, \mathrm{km}$ for $\nu_\tau$ detection. In the absence of detailed studies for alternatives at a neutrino factory, we adopt OPERA-like properties with 2~kt detector masses, as simulated in \Ref~\cite{Huber:2006wb} and described in \Ref~\cite{Autiero:2003fu} for the $\nu_e \rightarrow \nu_\tau$ channel at the neutrino factory. These detectors correspond to ND5 (see also \figu{ring}). Since there is no study for $\nu_\mu \rightarrow \nu_\tau$ yet, we follow the pragmatic approach in  \Ref~\cite{FernandezMartinez:2007ms} (see also \Ref~\cite{Donini:2008wz} for a similar approach) and use five times the signal and five times the backgrounds from \Ref~\cite{Autiero:2003fu}, because no charge identification is required, and therefore all tau decay modes (the hadronic ones as well) may be exploited at the price of higher backgrounds. We assume that similar efficiencies and backgrounds can be used for both the neutrino and antineutrino channels, and we add all events from $\nu_e \rightarrow \nu_\tau$ and $\bar\nu_\mu \rightarrow \bar\nu_\tau$ (same for the other polarity). For the detector geometry, we use OPERA-like measures, meaning that the detectors will perform as ND5 from the flux point of view. We show the differential decay rates corrected by the efficiency factors for these detectors in \figu{flux5}. Compared to the far distance limit, which is also shown for comparison, there is already a substantial modification of the fluxes. Since we do not measure the $\nu_\tau$ cross sections explicitely, we assume increased 30\% signal normalization errors. However, at the sensitivity limit for $\epsilon^s$, the signal normalization errors will be of secondary importance, meaning that the cross sections do not need to be exactly known for this application.
Note that this toy implementation of a $\nu_\tau$ detector may be in fact too primitive in practice, it should only serve as an example. For instance, the scanning load in an OPERA-like emulsion cloud chamber might be extremely high, which means that a different detector technology with at least similar efficiencies and vertex reconstruction abilities might be desirable (\eg, liquid argon). Maybe the $\nu_\mu$ and $\nu_\tau$ near detection can even be done at the same place with the same detectors.

\begin{table}[t]
\begin{center}
\begin{tabular}{lrr}
\hline
& Without $\nu_\tau$ ND5 & With $\nu_\tau$ ND5 \\
\hline
$|\epsilon^s_{e \tau}|$ & 0.004 & 0.0007\\
$|\epsilon^s_{\mu \tau}|$ & 0.4 & 0.0006 \\
$|\epsilon^m_{e \tau}|$ & 0.004 & 0.004 \\
$|\epsilon^m_{\mu \tau}|$ & 0.02 & 0.02 \\
\hline
\multicolumn{3}{l}{{\bf With correlation $\boldsymbol{\epsilon^s_{\mu \tau} = - (\epsilon^m_{\mu \tau})^*}$}} \\
$|\epsilon^s_{\mu \tau}|$,$|\epsilon^m_{\mu \tau}|$ & 0.003 & 0.0006 \\
\hline
\end{tabular}
\end{center}
\caption{\label{tab:nsi} Sensitivity to the NSI parameters without (left column) and with (right column) $\nu_\tau$ near detectors at the 90\% confidence level. For the non-diagonal NSI, the phases have been marginalized over.}
\end{table}

We give the sensitivities for the source NSI $|\epsilon^s_{e \tau}|$ and $|\epsilon^s_{\mu \tau}|$, and the matter NSI $|\epsilon^m_{e \tau}|$, and $|\epsilon^m_{\mu \tau}|$ in \Tab~\ref{tab:nsi} with and without $\nu_\tau$ near detectors. As expected, the source NSI improve significantly in the presence of the near $\nu_\tau$ detectors -- especially $|\epsilon^s_{\mu \tau}|$. Compared, for instance to the NOMAD results~\cite{Astier:2001yj} of 0.086
for $|\epsilon^s_{e \tau}|$ and 0.013 for $|\epsilon^s_{\mu \tau}|$ (90\% CL), the results can be improved by about two orders of magnitude.\footnote{There are also other theoretical limits, such as from rare lepton decays~\cite{Antusch:2006vwa}. However, the neutrino factory limits will be even stronger.} The bounds of the matter NSI $|\epsilon^m_{e \tau}|$ and $|\epsilon^m_{\mu \tau}|$ compare very well to the ones in \Ref~\cite{Kopp:2008ds} even for our different systematics treatment. Note that the
bound on $|\epsilon^m_{\mu \tau}|$ is rather weak if $\epsilon^m_{\mu \tau}$ is allowed to be complex. If, however, the NSI are assumed to originate from effective dimension six operators involving four lepton doublets, and charged lepton flavor violation is sufficiently suppressed, there is a general connection between source and matter NSI as~\cite{Gavela:2008ra}
\begin{equation}
\epsilon^s_{\mu \tau} = - (\epsilon^m_{\mu \tau})^* \, .
\label{equ:nsicorr}
\end{equation}
This means that the bounds on $|\epsilon^s_{\mu \tau}|$ applies, under this condition, to $|\epsilon^m_{\mu \tau}|$ as well, and vice versa. In particular, the bound for $|\epsilon^m_{\mu \tau}|$ becomes quite strong in this case, even better than the bound from lepton universality $1.9 \, 10^{-3}$ (90\% CL)~\cite{Antusch:2008tz} in the presence of the $\nu_\tau$ near detectors (see last row in \Tab~\ref{tab:nsi}).

An interesting application of the correlation in \equ{nsicorr} is the test of non-standard CP violation (NSI-CPV). Such NSI-CPV has been studied in \Refs~\cite{Gonzalez-Garcia:2001mp,FernandezMartinez:2007ms,Goswami:2008mi,Winter:2008eg,Altarelli:2008yr}.  Since the zero-distance effect is proportional to $| \epsilon^s |^2$, there is no sensitivity to CP violation in our near detectors. However, if the source and matter NSI are correlated through \equ{nsicorr}, the near detectors can be used to constrain $\epsilon^s_{\mu \tau}$, and the far detectors can be used to measure the NSI-CPV in $\epsilon^m_{\mu \tau}$ very well (as discussed in \Ref~\cite{Winter:2008eg}). In summary, our configuration can identify such NSI-CPV for $\phi^m_{\mu \tau}=\pi/2$ (maximal CPV) in that case down to $|\epsilon^m_{\mu \tau}|\simeq 0.0005$ (90\% CL),
whereas without $\nu_\tau$ near detectors it can only be measured down to 0.003.
This is comparable to the result in \Ref~\cite{FernandezMartinez:2007ms}, where the additional baseline $L=130 \, \mathrm{km}$ was proposed. Without the correlation in \equ{nsicorr}, the CPV in $\epsilon^s_{\mu \tau}$ can be measured down to 0.001 in the far detectors (for $\phi^s_{\mu \tau}=-\pi/2$), meaning that the additional contributions from $\epsilon^m_{\mu \tau}$ are be counter-productive in the presence of the correlation (see Fig.~1b in \Ref~\cite{Huber:2002bi}), and the  CPV in $\epsilon^m_{\mu \tau}$ down to 0.02 only, in consistency with \Ref~\cite{Winter:2008eg}.

\section{Summary and conclusions}
\label{sec:summary}

We have discussed the impact of near detectors on standard oscillation physics at a high and low energy neutrino factory, and the possible use of near detectors for $\nu_\tau$ appearance. We have included the geometry of the source and the detectors in the event rate calculations, and we have defined qualitatively different near detectors. For this definition, it has turned out to be useful to consider the near detector limit (detector captures whole beam flux) and the far detector limit (detector observes similar spectrum to that of a far detector), as well as intermediate cases between these limits. We have demonstrated that a neutrino factory near detector can effectively be parameterized as a far detector with an energy-dependent effective area. In this case, the decay straight can be treated as a point source with an effective baseline which is the geometric mean between the shortest and longest distance to the detector. 

In order to test the impact of the near detectors on physics, we have used the IDS-NF (International Design Study for the Neutrino Factory) baseline setup as a starting point. In the current baseline setup, the near detectors are not yet defined explicitely. Therefore, we have refined the systematics treatment including cross section errors fully uncorrelated among the energy bins, but fully correlated among all channels measuring muon neutrinos or antineutrinos. In addition, we have included flux normalization errors fully correlated among all energy bins, but uncorrelated among the different storage rings and muon polarities. As a conservative assumption, we have kept relatively large (uncorrelated) background uncertainties in all channels. We have then tested the impact of this systematics treatment and the different near detector types on standard oscillation physics, and we have considered $\nu_\tau$ appearance in near detectors for non-standard interactions.

For the characteristics of the near detectors for standard oscillation measurements, we have found that two near detectors should be operated on-axis (with respect to the decay straights) if the muons and anti-muons circulate in different directions in a racetrack-shaped ring. These two detectors measure the products between the (uncorrelated) flux normalizations and muon or muon anti-neutrino cross sections. For a two baseline operation of a high energy neutrino factory with two storage rings, we have demonstrated that near detectors in one storage ring might be sufficient. The near detectors should be at least as good as the far detectors in terms of efficiencies, energy resolution, \etc, but the technical details should be of secondary importance. The near detectors can be very small for the systematics considered as long as the event rates are similar to that of the far detectors -- a few kilograms are basically sufficient. In this case, the near detectors can be operated quite closely to the source in the ``far distance limit'', which means that the observed spectrum is similar to that of a far detector. For larger near detectors, the event rates are very high in all energy ranges so that a possibly different spectrum does not significantly affect the near-far extrapolation even in the most extreme cases. 

As far as the standard oscillation parameter measurements are concerned, the results depend on the number of far detectors used. If the neutrino factory has only one baseline, such as a high energy ($E_\mu=25 \, \mathrm{GeV}$) neutrino factory with $L=4000 \, \mathrm{km}$ or a low energy ($E_\mu=4.12 \, \mathrm{GeV}$) neutrino factory with $L=1290 \, \mathrm{km}$, the near detectors are mandatory for the leading atmospheric parameter measurements. For the $\theta_{13}$ and mass hierarchy discovery reaches, we have not found any significant impact, because these measurements are background limited. For the CP violation discovery reach, the near detectors are important, because the unknown atmospheric parameters lead to intrinsic unknown backgrounds from the CP-even terms, which can be controlled by the near detectors. For small $\stheta \lesssim 10^{-2}$, this effect hardly depends on the flux knowledge, which means that flux monitoring is of secondary interest. For large $\stheta$, however, fluxes monitored to the level of 0.1\% help to understand the measurement. In particular, it is ensured that the CP violating effect does not come from a wrongly assumed neutrino-antineutrino cross section ratio.

For a two baseline high energy neutrino factory, such as the IDS-NF baseline setup with $L_1=4000 \, \mathrm{km}$ combined with $L_2=7500 \, \mathrm{km}$, we have demonstrated that the considered systematical errors cancel even without near detectors and good flux monitoring. In such a setup, the same product of fluxes and cross sections is measured in both far detectors. Therefore, a two baseline neutrino factory should be very robust with respect to systematics, no matter how many near detectors are used. Of course, in staging scenarios where only one baseline is operated first, near detectors are still required, as well as they may be needed for different purposes.

Because of the particular correlations of the cross section and flux normalization errors, our high energy neutrino factory performs slightly better than the current IDS-NF baseline version, where all signal and background normalization errors are fully uncorrelated. Therefore, systematics may be actually quite conservative for the IDS-NF baseline. Note that we have not even considered a background extrapolation from the near to far detector (depending on the properties of the near detectors), but instead adopted the conservative IDS-NF uncorrelated errors. However, we have not considered energy calibration errors, where the near detectors may also help, and detector-dependent systematical uncertainties, such as errors in the detection efficiencies, which may partially cancel if similar far detectors are used. Therefore, a more realistic systematics treatment may produce results somewhere in the middle. For the low energy neutrino factory, our systematics treatment reproduces the reference setup with effective near detector simulations very well.
\begin{figure}[t]
\begin{center}
\includegraphics[width=\textwidth]{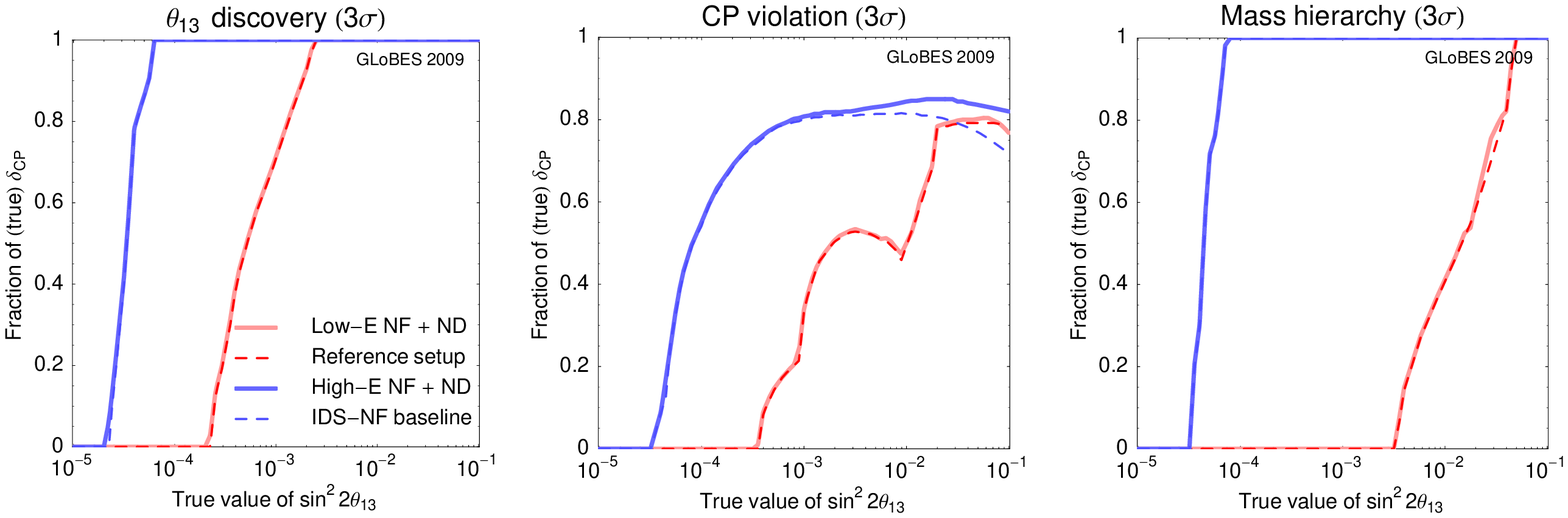}
\end{center}
\mycaption{\label{fig:lehe} $\theta_{13}$, CP violation, and mass hierarchy discovery reaches ($3 \sigma$) as a function of the fraction of $\deltacp$ and (true) $\stheta$. The solid curves represent our near detector-far detector simulation for the high (left curves) and low energy (right curves) neutrino factories, whereas the dashed curves represent the IDS-NF~\cite{ids} and Bross et al~\cite{Bross:2007ts} reference setups.}
\end{figure}
We summarize our main results in \figu{lehe}, where the $\theta_{13}$, CP violation, and mass hierarchy discovery reaches are shown for both the IDS-NF and low energy neutrino factory. The solid curves represent our near detector-far detector simulation for the high (left curves) and low energy (right curves) neutrino factories, whereas the dashed curves represent the reference setups. This figure illustrates the impact of the different systematics, and it compares, for the first time, the low and high energy neutrino factory options for the same input assumptions.

Finally, we have considered near detectors with OPERA-like properties for $\nu_\tau$ appearance to measure non-standard interactions (NSI).  We have demonstrated the NOMAD limits for the zero-distance effect could be improved by about two orders of magnitude. In addition, we have discussed the theoretical possibility that source and matter NSI are connected at a generic level assuming dimension six effective operators. In this case, even the matter NSI parameter $|\epsilon^m_{\mu \tau}|$ becomes quite strongly limited, exceeding the bound from lepton universality. Furthermore, CP violation from matter NSI may become measurable down to 0.0005 in $|\epsilon^m_{\mu \tau}|$.

We conclude that two near detectors are mandatory for a successful neutrino factory operation with one baseline, and a good enough flux monitoring will be useful for some physics measurements. However, a two baseline high energy neutrino factory may prove to be very robust with respect to systematical errors, even if the systematics goals in the initial one baseline operation phase cannot be achieved. In either case, near detectors will be mandatory to constrain certain new physics effects. Since the statistics is typically very high, the detailed requirements for the near detectors will be most likely driven by flux monitoring and new physics measurements, rather than the cross section uncertainties.

\subsubsection*{Acknowledgments}

We would like to thank Steve Geer and the members of the IDS-NF for useful discussions. JT is indebted to Toshihiko Ota for comments on the flux normalization.
This work has been supported by the Emmy Noether program of DFG under contract WI 2639/2-1.

\begin{appendix}
\section{Details of the systematics treatment}
\label{app:syst}

For the systematics treatment, we use the pull method.
As an example, we describe here the systematics treatment of the high energy neutrino factory.
The total $\chi^2$ is given by
\begin{equation}
\chi^2 = \sum\limits_{C=1}^{4} \chi^2_{F1,C} + \sum\limits_{C=1}^{4} \chi^2_{F2,C} + \chi^2_{N1} + \chi^2_{N2} + \chi^2_{\mathrm{Pull}} \, ,
\end{equation}
summed over the $\chi^2_{D, C}$ of detector $D$ and channel $C$. For the detectors, we have the far detector $F1$ at the $4 \, 000 \, \mathrm{km}$ baseline, the far detector $F2$ at the $7 \, 500 \, \mathrm{km}$ baseline, the near detector $N1$ behind the $\mu^+$ decay section of the storage ring $S1$ to $F1$, and the near detector $N2$ behind the $\mu^-$ decay section of the storage ring $S1$ to $F1$ (\cf, \figu{ring}).
For the far detector channels $C$, we have $\nu_\mu$ appearance (1), $\bar\nu_\mu$ appearance (2), $\nu_\mu$ disappearance (3), and $\bar\nu_\mu$ disappearance (4), whereas the near detectors only measure un-oscillated $\nu_\mu$ and $\bar\nu_\mu$, respectively.

Let the theoretical (fit) rate $T^i$ be the rate in the $i$th bin, which is composed of the signal rate $S^i$ and the background rate $B^i$. Then we have for the different fit rates in the bins using the $a$'s as the auxiliary parameters of the pull method to be marginalized over\footnote{Here we have simplified the background treatment a bit. In fact, we include the bin-dependent cross section uncertainties in the backgrounds from charge mis-identification. However, there is hardly any effect of this more refined treatment.}
\begin{eqnarray}
T_{F1,1}^i & = & (1 + a^{\mu^+, \, S1}_{\mathrm{Flux}} + a^{\nu_\mu, \, i}_{\mathrm{Xsec}}  ) \, S_{F1,1}^i +  (1 + a^1_{\mathrm{BG}}  ) \, B_{F1,1}^i \, , \\
T_{F1,2}^i & = & (1 + a^{\mu^-, \, S1}_{\mathrm{Flux}} + a^{\bar\nu_\mu, \, i}_{\mathrm{Xsec}}  ) \, S_{F1,2}^i +  (1 + a^2_{\mathrm{BG}}  ) \, B_{F1,2}^i \, , \\
T_{F1,3}^i & = & (1 + a^{\mu^-, \, S1}_{\mathrm{Flux}} + a^{\nu_\mu, \, i}_{\mathrm{Xsec}}  ) \, S_{F1,3}^i +  (1 + a^3_{\mathrm{BG}}  ) \, B_{F1,3}^i \, , \\
T_{F1,4}^i & = & (1 + a^{\mu^+, \, S1}_{\mathrm{Flux}} + a^{\bar\nu_\mu, \, i}_{\mathrm{Xsec}}  ) \, S_{F1,4}^i +  (1 + a^4_{\mathrm{BG}}  ) \, B_{F1,4}^i \, , \\
T_{F2,1}^i & = & (1 + a^{\mu^+, \, S2}_{\mathrm{Flux}} + a^{\nu_\mu, \, i}_{\mathrm{Xsec}}  ) \, S_{F2,1}^i +  (1 + a^5_{\mathrm{BG}}  ) \, B_{F2,1}^i \, , \\
T_{F2,2}^i & = & (1 + a^{\mu^-, \, S2}_{\mathrm{Flux}} + a^{\bar\nu_\mu, \, i}_{\mathrm{Xsec}}  ) \, S_{F2,2}^i +  (1 + a^6_{\mathrm{BG}}  ) \, B_{F2,2}^i \, , \\
T_{F2,3}^i & = & (1 + a^{\mu^-, \, S2}_{\mathrm{Flux}} + a^{\nu_\mu, \, i}_{\mathrm{Xsec}}  ) \, S_{F2,3}^i +  (1 + a^7_{\mathrm{BG}}  ) \, B_{F2,3}^i \, , \\
T_{F2,4}^i & = & (1 + a^{\mu^+, \, S2}_{\mathrm{Flux}} + a^{\bar\nu_\mu, \, i}_{\mathrm{Xsec}}  ) \, S_{F2,4}^i +  (1 + a^8_{\mathrm{BG}}  ) \, B_{F2,4}^i \, , \\
T_{N1}^i & = & (1 + a^{\mu^+, \, S1}_{\mathrm{Flux}} + a^{\bar\nu_\mu, \, i}_{\mathrm{Xsec}}  ) \, S_{N1}^i +  (1 + a^9_{\mathrm{BG}}  ) \, B_{N1}^i \, , \\
T_{N2}^i & = & (1 + a^{\mu^-, \, S1}_{\mathrm{Flux}} + a^{\nu_\mu, \, i}_{\mathrm{Xsec}}  ) \, S_{N2}^i +  (1 + a^{10}_{\mathrm{BG}}  ) \, B_{N2}^i \, . 
\end{eqnarray} 
The auxiliary parameters describe the following types of systematics errors:  $a^{\mu^\pm, \, Sk}_{\mathrm{Flux}}$ are flux normalization errors for storage ring $Sk$ and $\mu^\pm$ stored (fully correlated among all bins and detectors in the same beam), $a^{\nu_\mu, \, i}_{\mathrm{Xsec}}$ represent the neutrino cross section errors (fully correlated among all channels measuring $\nu_\mu$, but uncorrelated among the bins),  $a^{\bar\nu_\mu, \, i}_{\mathrm{Xsec}}$ represent the antineutrino cross section errors (fully correlated among all channels measuring $\bar\nu_\mu$, but uncorrelated among the bins), and 
$a^k_{\mathrm{BG}}$ the background errors (fully uncorrelated among all channels, but fully correlated among all bins). Altogether, there are 32~cross section errors, ten background normalization errors, and four flux normalization errors, \ie, 46~systematical errors in total.

The $\chi^2$ for each detector or channel is then obtained as
\begin{equation}
\chi^2_{D,C} = \sum_{i=1}^{16}  2 \left[ T^i_{D, C} - O^i_{D, C} + O^i_{D, C} \cdot \mathrm{ln} \left( \frac{O^i_{D, C}}{T^i_{D, C}} \right)  \right] \, ,
\end{equation}
where the Poissonian $\chi^2$,
in which $O$ is the observed/simulated rate and $T$ is the theoretical/fit rate, is used (near detectors analogously). In addition, the pull $\chi^2$ is given by
\begin{equation}
\chi^2_{\mathrm{Pull}} =  \sum\limits_{k=1}^2 \left( \frac{ a^{\mu^+, \, Sk}_{\mathrm{Flux}}}{\sigma_{\mathrm{Flux}}} \right)^2 + \sum\limits_{k=1}^2 \left( \frac{ a^{\mu^-, \, Sk}_{\mathrm{Flux}}}{\sigma_{\mathrm{Flux}}} \right)^2+  \sum\limits_{i=1}^{16} \left(  \frac{a^{\nu_\mu, \, i}_{\mathrm{Xsec}}}{\sigma_{\mathrm{Xsec}}} \right)^2+  \sum\limits_{i=1}^{16} \left(  \frac{a^{\bar\nu_\mu, \, i}_{\mathrm{Xsec}}}{\sigma_{\mathrm{Xsec}}} \right)^2+ \sum\limits_{k=1}^{10} \left( \frac{a^k_{\mathrm{BG}}}{\sigma_{\mathrm{BG}}} \right)^2 \, .
\end{equation}
For the high energy neutrino factory, we (conservatively) assume $\sigma_{\mathrm{Flux}}=2.5\%$, $\sigma_{\mathrm{Xsec}}=30\%$, and  $\sigma_{\mathrm{BG}}=20\%$, unless explicitely stated otherwise.
\end{appendix}

\end{document}